\documentclass{aa}
\usepackage{booktabs}
\usepackage{graphicx}
\usepackage{comment}
\usepackage{appendix}
\usepackage{enumitem}
\usepackage[breaklinks=true]{hyperref}
\usepackage{url}
\usepackage{txfonts}
\usepackage{xcolor}
\begin{document}

   \title{Evidence of flare and coronal mass ejection-driven shock acceleration of relativistic protons on 21 November 2024}

   \titlerunning{Flare and CME-shock acceleration on 21 November 2024}

   \subtitle{}

   \author{M. Jarry\inst{1}
          \and
          A. Papaioannou\inst{1}
          \and
          G. Vasalos\inst{1}
          \and
          N. Talebpour Sheshvan\inst{2}
          \and
          D. E. Morosan\inst{3}
          \and
          A. Kouloumvakos\inst{4}
          \and
          G. M. Mason\inst{4}
          \and
          E. Lavasa\inst{1}
          \and
          A. P. Rouillard\inst{2}
          \and
          A. Anastasiadis\inst{1}
          \and
          B. Heber\inst{5}
          \and
          M. H\"orl\"ock\inst{5}
          }

   \institute{
         Institute for Astronomy, Astrophysics, Space Applications and Remote Sensing (IAASARS), National Observatory of Athens, I. Metaxa \& Vas. Pavlou St., 15236 Penteli, Greece\\ \email{mjarry@noa.gr}
         \and
         Institut de Recherche en Astrophysique et Plan{\'e}tologie (IRAP), CNRS, Universit{\'e} de Toulouse, France
         \and
         Department of Physics and Astronomy, University of Turku, 20500 Turku, Finland
         \and
         Applied Physics Laboratory, Johns Hopkins University, Laurel, MD 20723, USA
         \and
         Institut für Experimentelle und Angewandte Physik, Christian-Albrechts-Universität zu Kiel, 24118 Kiel, Germany
         }

   \date{Received 30 October 2026 / Accepted ...}

 
  \abstract
   {On 21 November 2024 at 17:55 UTC, a solar flare erupted from far-side active region (AR) 13892, with enhanced hard X-ray flux measured by Solar Orbiter (SO).
   It was associated with a coronal mass ejection (CME) travelling at 1436 km s$^{-1}$ along the limb.
   This event produced a strong and widespread solar energetic particle (SEP) event at Earth, STEREO-A and SO, with proton energies exceeding 100 MeV.
   }
   {We investigate the relativistic particle acceleration during this event, focusing on separating flare- and shock-related contributions and linking spectral evolution to the CME-driven shock properties.
   }
   {We combine multi-spacecraft energetic-particle observations with velocity dispersion analysis (VDA), three-dimensional CME-driven shock reconstruction, and magnetic connectivity modelling. Radio observations and SEP composition provided independent diagnostics of shock- and flare-associated particle acceleration, respectively.
   }
   {VDA yields solar release times (SRTs) consistent with an early flare-accelerated particle population at all three observers, despite their broad range of helio-longitudes.
   At STEREO-A, the early release is followed immediately by magnetic connection to an efficient CME-driven shock.
   A 3D reconstruction of the type II radio burst location confirms a connection between Earth and an efficient shock-acceleration region several minutes later.
   Within the adopted models, the SO release precedes magnetic connection to the reconstructed front, disfavouring acceleration at the connected shock as the source of the earliest particles.
   These differences show that the subsequent shock contribution depends on magnetic connectivity and local shock efficiency.
   Despite this diversity, STEREO-A and SO show similar spectral evolution, while the near-Earth spectrum extends to relativistic energies and exhibits a high-energy rollover better reproduced by curved than single power-law spectra. SEP abundances also show signatures of flare-related acceleration at all three observers.
   }
   {The analysis suggests multiple acceleration processes acting non-uniformly across observers.
   The spectral evolution reflects their combined effects and subsequent interplanetary transport.
   }


   \maketitle

\section{Introduction}
\label{sect:intro}

Solar energetic particles (SEPs) are accelerated during solar eruptive events and can reach energies extending to several hundred of MeV or more. These particles are commonly associated with solar flares and coronal mass ejections (CMEs), and their acceleration is generally attributed either to magnetic reconnection processes occurring during flares or to shocks driven by fast CMEs propagating through the corona and heliosphere \citep[e.g.][]{Reames_1999, Desai_2016, Klein_Dalla_2017, Papaioannou_2025b}. Determining the dominant acceleration mechanism and identifying the location of the acceleration region remain major challenges in heliophysics.

The most energetic SEP events can produce relativistic particles detectable at ground level by neutron monitors \citep[][]{Gopalswamy_2012}. Such events mark extreme end of the SEP energy spectrum and provide valuable constraints on particle acceleration processes.
However, establishing the origin of these particles remains difficult because the timing and spatial evolution of solar eruptions involve complex interactions between flares, CME-driven shocks, and the large-scale coronal magnetic field \citep[see e.g.][]{Rouillard_2016, Klein_2022}.
In this context, several key questions remain open. In particular, it is still unclear whether the timing of SEP injections is consistent with acceleration by CME-driven shocks or whether flare-related processes may contribute significantly. 
Another important issue concerns the location of the acceleration region and its evolution as the eruption expands through the corona \citep{Dresing_2022}.
Finally, the magnetic connectivity between the acceleration region and the observing spacecraft plays a crucial role in determining the observed particle intensities and spectral characteristics \citep[e.g.][]{Rouillard_2012, Plotnikov_2017, RodriguezGarcia_2025, Dresing_2023, Jarry_2024}.

In this work, we investigate the solar event of the 21 November 2024, during which particles reached relativistic energies, using a combination of remote sensing observations, in situ particle measurements from multiple spacecraft, and three-dimensional reconstruction techniques.
The geometry of the expanding coronal shock is derived from imaging observations, while magnetic field lines connecting the Sun to the observing spacecraft are obtained using different magnetic field models.
By combining the reconstructed shock geometry with a global magnetohydrodynamic simulation of the corona, we derive the evolution of the shock parameters along the magnetic field lines connected to the observers.
This approach allows us to compare the timing and properties of the observed SEP event with the characteristics of the associated solar eruption.

This paper is organized as follows.
Section \ref{sect:observations} presents an overview of the observations associated with the solar eruption and the SEP event of the 21 November 2024, including both remote sensing observations and in situ particle measurements.
In Section \ref{sect:3D_reconstrcution}, we reconstruct the three-dimensional geometry of the coronal shock and the magnetic field lines connecting the Sun to the observing spacecraft, and derive the shock parameters along these field lines.
Section \ref{sect:acceleration_analysis} investigates the relationship between the solar activity and the observed SEP event by comparing particle timings, radio emissions, and the temporal evolution of SEP intensities and spectra.
Finally, the results are discussed and summarized in Section \ref{sect:discuss_conclusion}.

\section{Observations}
\label{sect:observations}

On 21 November 2024, a strong solar energetic particle (SEP) event was associated with an eruption located on the far side of the Sun as viewed from Earth.
This section provides an overview of the observations associated with this event.
We first present the remote sensing observations of the solar eruption, including X-ray, EUV, and radio signatures, as well as the identification of the source active region and the associated coronal mass ejection.
We then describe the in situ measurements of solar energetic particles detected by multiple spacecraft, including the determination of particle onset times and velocity dispersion analyses.
Finally, we investigate the energy spectra of the energetic particles measured at different heliospheric locations.

\subsection{Remote sensing observations}
\label{subsect:remote_obs}

Due to its location, the flare did not produce any major soft X-ray (SXR) signature in the Geostationary Operational Environmental Satellite \citep[GOES;][]{Hanser_1996} dataset, which remained in the C-class range, starting at 17:52 UT and peaking at 18:07 with a maximum of C6.0, visible in panel (f) of Fig.~\ref{fig:SEP_and_obs} (red curve).
The Spectrometer/Telescope for Imaging X-rays \citep[STIX;][]{Krucker_2020} onboard Solar Orbiter \citep[SO;][]{Muller_2013}, which was also positioned on the far side relative to the parent active region, recorded an increase in the 15--25~keV range, starting around 17:55 and peaking at 18:01, with an estimated GOES class of C4.
Despite their low amplitude, these X-rays flares could be measured because of the very low background in the instruments.

\begin{figure}[t]
\centering
\includegraphics[width=0.43\columnwidth]{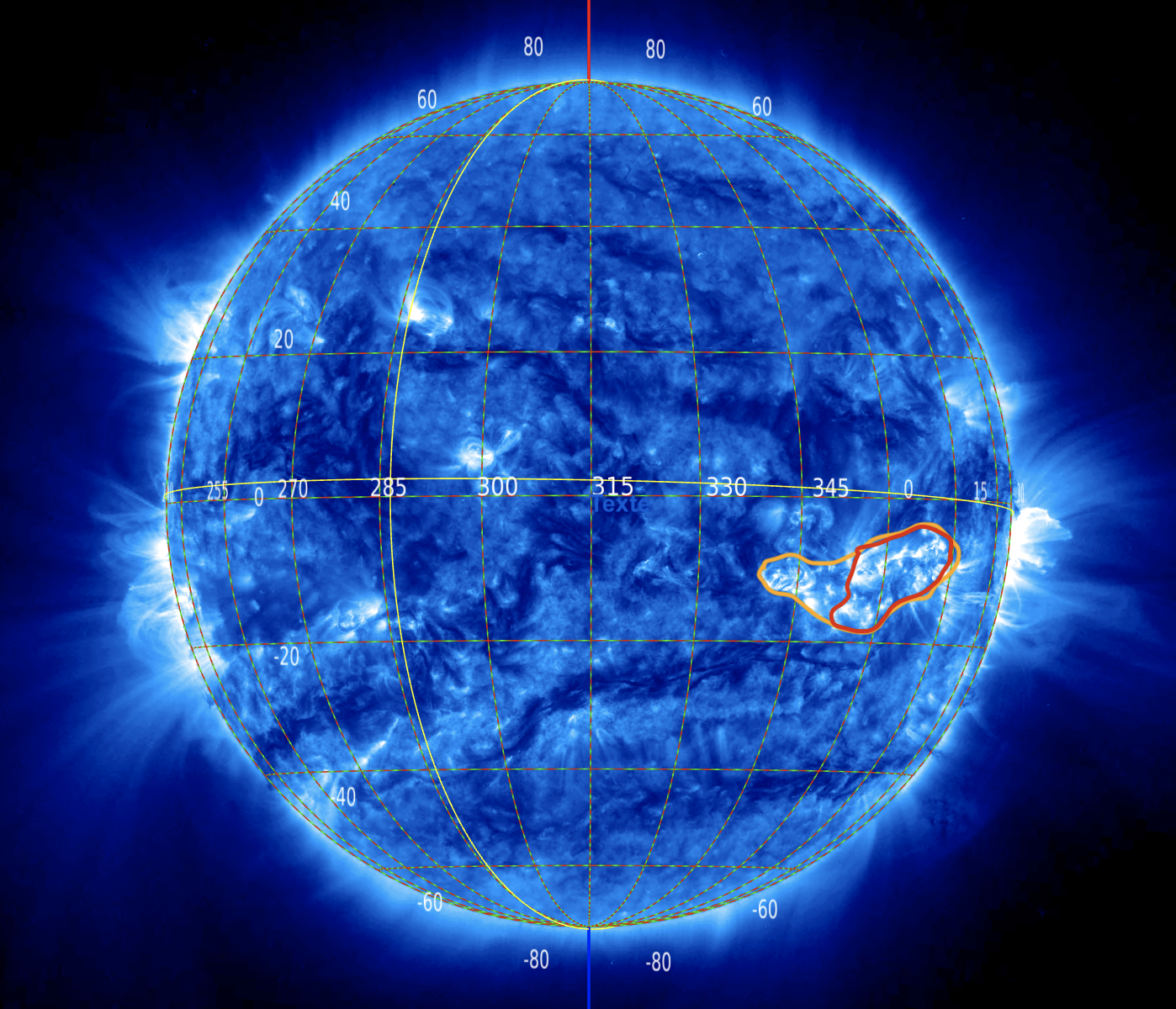}
\includegraphics[width=0.47\columnwidth]{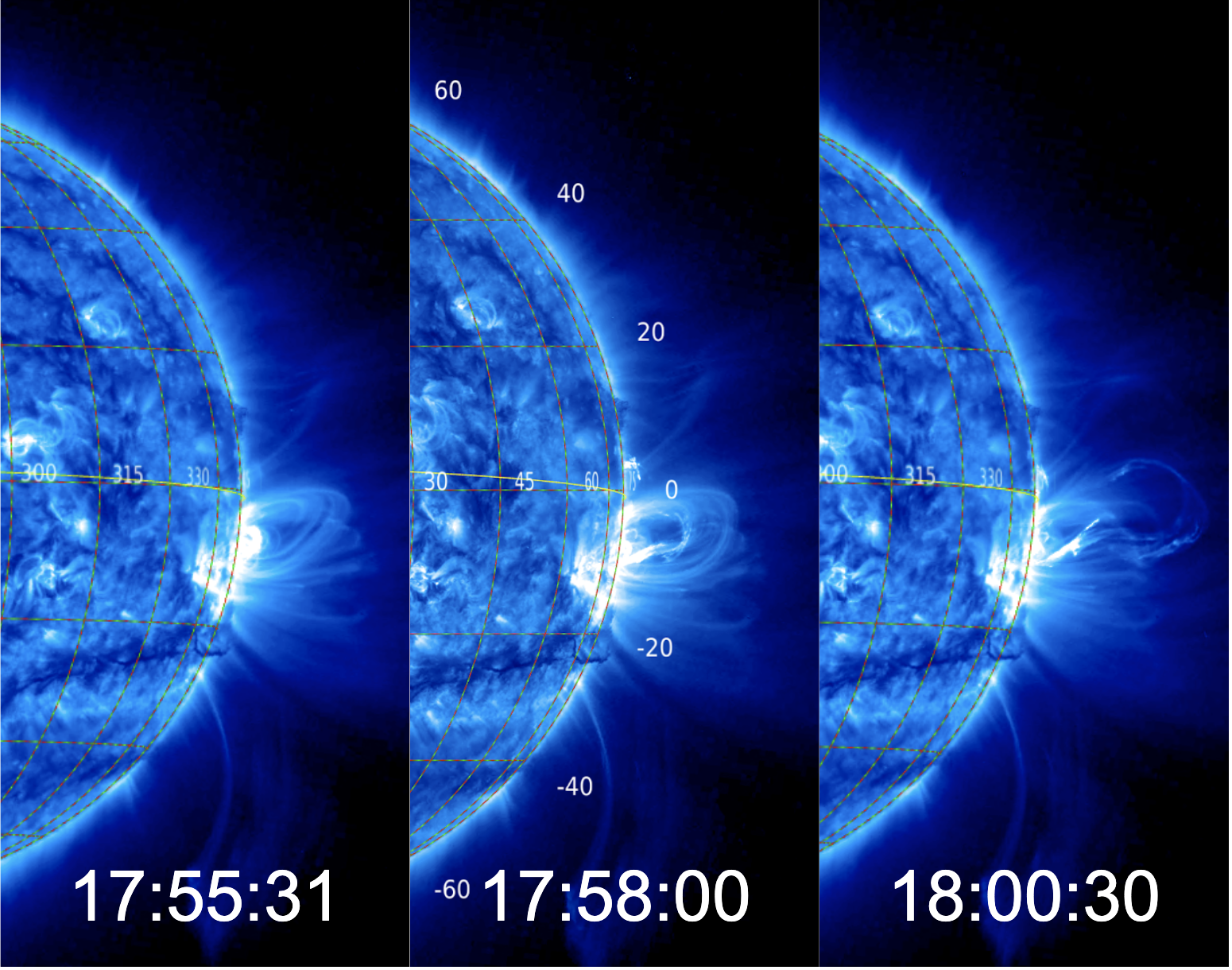}
\caption{STEREO-A EUVI images prior and at the time of the flare eruption, i.e 17:55:31 UT on the 21 November 2024. Left panel: identification of the responsible AR~13892 with the red line (extended until the orange line) on the 18 November 2024 at 17:20:31 UT. Right panel: three images following each other showing the first flare light and the extension of the post-flare loop.} 
\label{fig:EUVI_AR_eruption}
\end{figure}

\begin{figure}[t]
\centering
\includegraphics[width=\columnwidth]{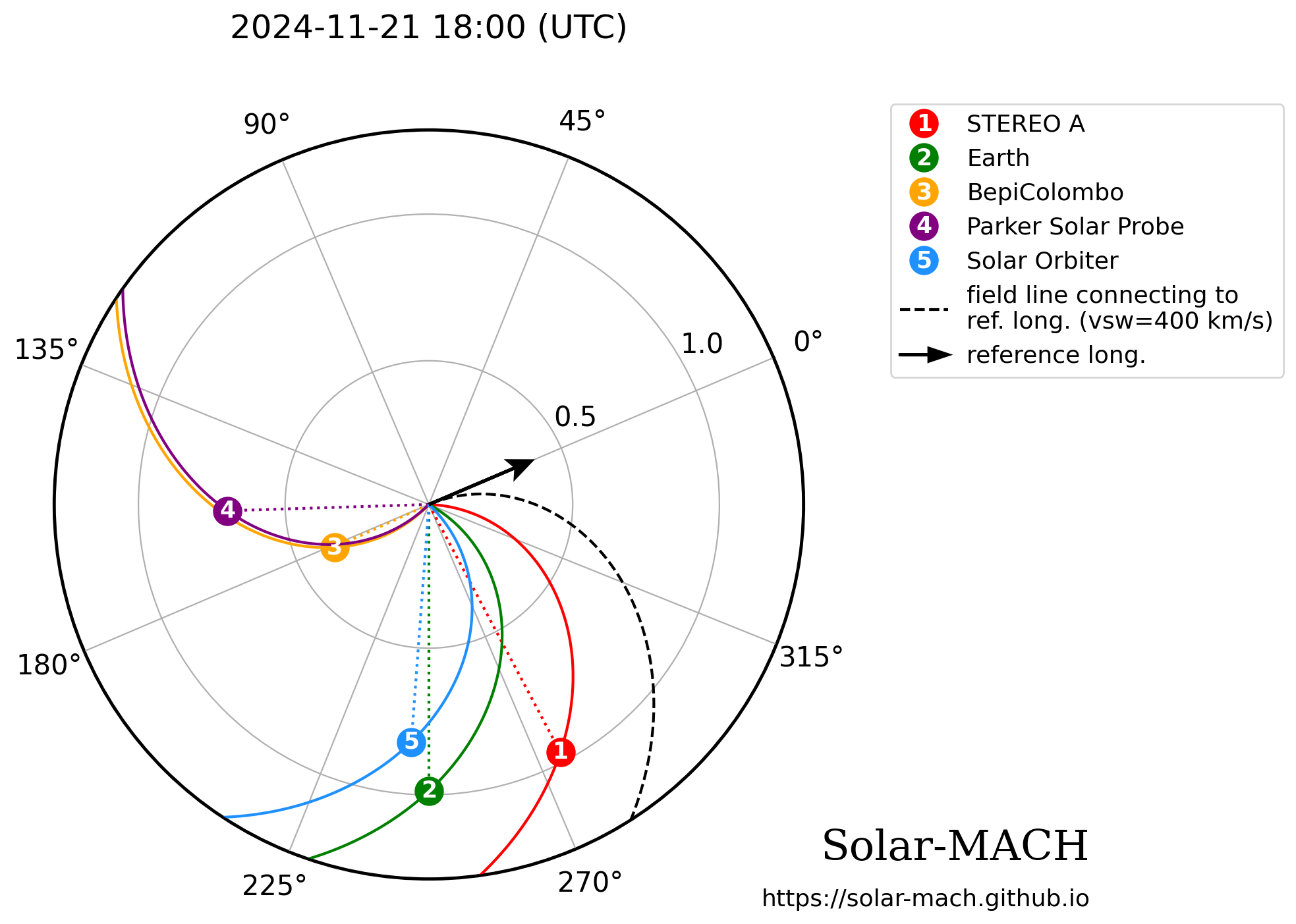}
\caption{Multi-spacecraft longitudinal configuration and magnetic connectivity at 18:00 UT on 21 November 2024. The black arrow indicates the position of the associated solar flare, while the dashed spiral represents the nominal magnetic field line connecting to the flare site. The plot was produced using the Solar-MACH online tool\citep{Gieseler_2023}.}  
\label{fig:sol_MACH}
\end{figure}

The flare location reported by STIX at S09E30\footnote{\url{https://github.com/hayesla/stix_flarelist_science/tree/main}} and by SolarSoft at S08E31\footnote{\url{https://www.lmsal.com/solarsoft/latest_events_archive/events_summary/2024/11/21/gev_20241121_1752/index.html}}, in Stonyhurst coordinates, i.e attached to the active region (AR)~13901, very active before the main event.
Exploring this further, we find that the source region is at NOAA AR~13892, identified using the Solar-Terrestrial Relations Observatory \citep[STEREO-A;][]{Kaiser_2008} EUV observations, and located between longitudes = [3$^{\circ}$, 13.5$^{\circ}$] and latitudes = [-13$^{\circ}$, -5$^{\circ}$] in Carrington coordinates (respectively CRLN and CRLT). This AR is identified using a red line in left panel of Fig.~\ref{fig:EUVI_AR_eruption}.
The first flare light was estimated around $(6 \pm 2)^{\circ}$ CRLN and $(-8 \pm 2)^{\circ}$ CRLT at approximately 17:55:00 $\pm$ 00:02:30~UT in STA/EUVI, and the expanding flare loop is clearly visible in right panel of Fig.~\ref{fig:EUVI_AR_eruption}.
The Sun presents a lot of other active zones, including neighbouring AR~13893 and AR~13898 ($\sim$S17W40) that are very well magnetically connected to Earth and SO spacecraft (see Section \ref{subsect:mag_connectivity}), but also the previously mentioned AR~13901, initially identified as the responsible region because the real one was far side, and a zone around S07W64 (CRLN $\sim312$). A complete description of previous solar activity is descried in Appendix \ref{annex: solar_activity}.

A CME followed shortly after the flare and observed at the west solar limb as viewed from Earth.
The CME was wide and subsequently appeared as a halo in the coronagraphs.
The CME first appearance was in SOHO/LASCO/C2 coronograph at 18:12 UT, and its height–time evolution is represented in Fig.~\ref{fig:SEP_and_obs}f (black points).
The blue line represents the linear fit to these measurements, yielding a CME speed of $\sim$1436 km s$^{-1}$, but it is that the 2nd order fit is more adapted to the measurements (see Fig.~\ref{fig:SOHO-LASCO_CME_fit} in Appendix \ref{annex: additional_material}).
This one give a speed value of 1217 km/s at 20 solar radius and 980 km/s at final height, i.e 24.6 solar radius.
Panel (g) of Fig.~\ref{fig:SEP_and_obs} presents the radio spectrogram recorded by STEREO/WAVES (SWAVES), showing type III radio bursts beginning at 17:55 UT and a type II burst starting at 18:10 UT.
Although Wind/WAVES RAD1 data contain a gap during this interval, observations from RAD2 confirm the onset of the type III burst at 17:55 UT.
Sub-section \ref{subsect:radio_burst} is dedicated to the radio data analysis, including the 2D and 3D reconstruction of the type II burst.

At the time of the eruption, multiple spacecraft were distributed over a wide range of helio-longitudes and heliocentric distances, providing comprehensive coverage of the event.
Figure~\ref{fig:sol_MACH} shows from a northern point of view the heliosphere and magnetic field lines connecting each satellite/object to the Sun using the Parker spiral model, in Carrington coordinates.
Table \ref{tab:fps_and_flare_locations} summarises the spacecraft locations, distances and Parker spiral footpoints.
STEREO-A (STA) was located at (CRLN, CRLT) = $(274.9^{\circ}, -1.6^{\circ})$, corresponding to the most favourable nominal magnetic connection to the parent active region.
Near-Earth spacecraft were located at (CRLN, CRLT) = $(246.9^{\circ}, 2.0^{\circ})$, while Solar Orbiter was positioned at $(242.7^{\circ}, 6.3^{\circ})$.
Further eastward, BepiColombo (Bepi) and Parker Solar Probe \citep[PSP;][]{Fox_2016} were respectively located at (CRLN, CRLT) = $(181.4^{\circ}, 3.6^{\circ})$ and (CRLN, CRLT) = $(158.7^{\circ}, 3.8^{\circ})$, at heliocentric distances of 0.70~au and 0.36~au, respectively.
These two observers were nominally connected to nearly the same Parker spiral magnetic field line, as discussed in Sect.~\ref{subsect:mag_connectivity}.
The black arrow represents the direction of propagation of the CME based on the flare location.

\subsection{Energetic particle observations}
\label{subsect:sep_obs}

\begin{figure}[t]
\centering
\includegraphics[width=0.99\columnwidth]{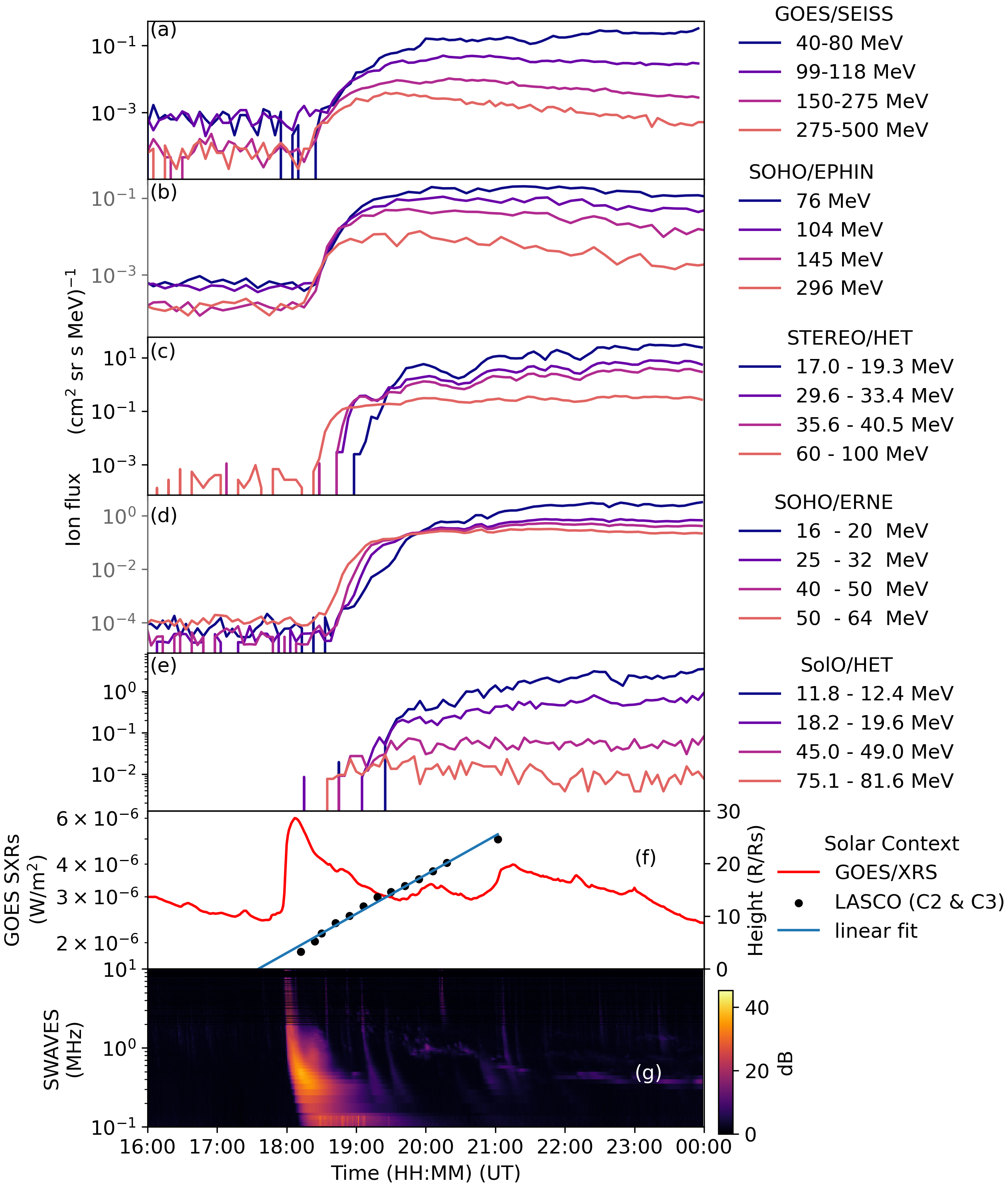}
\caption{Multi-spacecraft observations of the energetic particle event and associated solar activity on [date]. Panels (a)–(e) show ion flux time profiles measured at different heliospheric locations (see Fig.~\ref{fig:sol_MACH}) and energy ranges. Panel (a) presents GOES/SEISS observations. Panel (b) shows SOHO/EPHIN data. Panel (c) shows measurements from STEREO/HET, while panel (d) displays SOHO/ERNE proton fluxes. Panel (e) presents observations from SO/HET. In each panel, coloured curves correspond to different energy channels as indicated in the legend, and fluxes are given in units of (cm$^{2}$ sr s MeV)$^{-1}$
Panel (f) shows the SXR flux measured by GOES/XRS (red curve), together with the height–time evolution of the associated CME derived from LASCO C2 and C3 coronagraph observations (black points). The blue line indicates a linear fit to the CME height measurements. Panel (g) shows the radio dynamic spectrum recorded by SWAVES.}  
\label{fig:SEP_and_obs}
\end{figure}

We use energetic particle observations from the Solar and Heliospheric Observatory \citep[SOHO;][]{Domingo_1995}/Energetic and Relativistic Nuclei and Electron experiment \citep[ERNE;][]{Torsti_1995} and the Electron Proton Helium Instrument \citep[EPHIN;][]{Mueller-Mellin_1995}, STA/High Energy Telescope \citep[HET;][]{vonRosenvinge_2008}, GOES/Space Environment In-Situ Suite \citep[SEISS;][]{Krimchansky_2004, Kress_2021}, and SO/High Energy Telescope \citep[HET;][]{Rodriguez-Pacheco_2020}.
Figure~\ref{fig:SEP_and_obs} presents SEP measurements from these instruments, showing (from top to bottom): GOES/SEISS protons from 40–500 MeV (panel (a)), SOHO/EPHIN protons from 76-296 MeV (panel (b)), STA/HET covering 17–100 MeV (panel (c)), SOHO/ERNE spanning 16–64 MeV (panel (d)) and SO/HET observing particles in the $\sim$11-81 MeV range (panel (e)). PSP being completely disconnected from the source region seems not to have recorded any enhancement at this time.

As evident from spacecraft observations (Fig.~\ref{fig:SEP_and_obs}a,b), relativistic protons were detected in interplanetary space by SOHO/EPHIN (296 MeV) and by GOES/SEISS (E = 275–500 MeV).
The same population of relativistic particles (i.e. up to about 500 MeV) was also observed at Earth, where ground-based neutron monitors recorded a maximum increase of about $\sim$7\% in the count rates at the DOMC station on the Antarctic plateau\footnote{\url{https://gle.oulu.fi/}}.
These combined energetic particle observations led to the classification of this SEP event as the third ground-level enhancement (GLE) \citep{Papaioannou_macau_2022} of solar cycle 25, GLE76 - following the relativistic SEP events of May \citep{Papaioannou_2025a} and June 2024 \citep{Poluianov_2025}.

At STA/HET, the SEP event exhibits a prompt onset across all energy channels and a clear velocity dispersion in proton arrival times.
This is illustrated in Figure~\ref{fig:SEP_and_obs}c, which shows the time profiles of the highest-energy channel (E = 60–100 MeV) together with four additional representative channels, spanning from 17-40 MeV.
The SEP event observed by STA/HET was intense and long-lasting, with proton energies reaching more than 80 MeV and remaining above background levels for more than two days.
A key factor contributing to the high intensities recorded at STA was the spacecraft’s close magnetic connection to both the source active region (within 11-28°, see Fig.~\ref{fig:magnetogram} and Table~\ref{tab:fps_and_flare_locations}) and the propagating CME-driven shock, as discussed in Section \ref{subsect:MHD_parameters}.
See Sect. \ref{sect:acceleration_analysis} for a complete analysis of SEP acceleration during this event.

\begin{figure}[t]
\centering
\includegraphics[width=0.7\columnwidth]{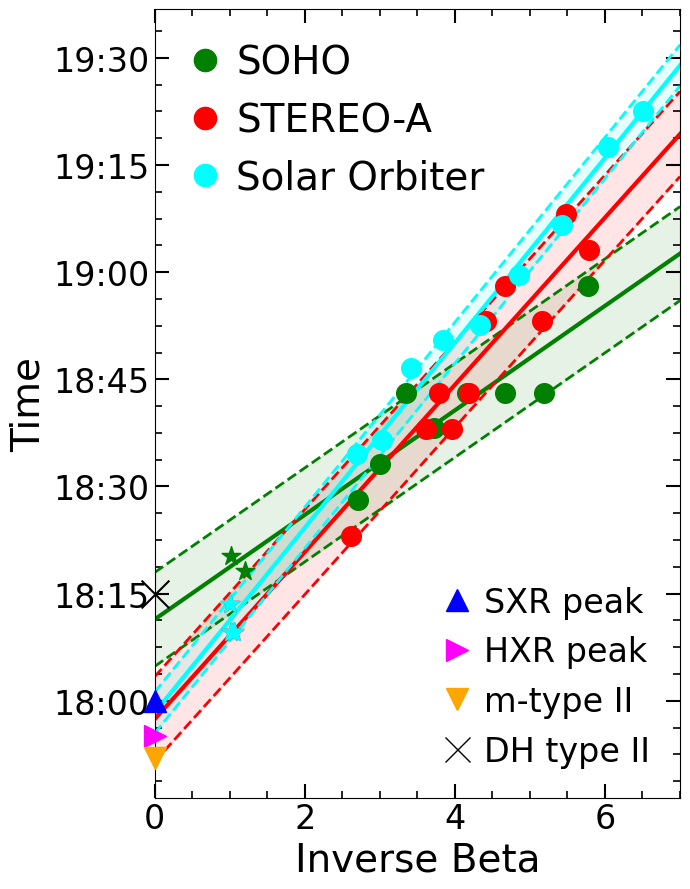}
\caption{VDA results for STA/HET protons (red filled circles) and electrons (red filled stars), SOHO/ERNE (green filled circles) and SOHO/EPHIN electrons (green filled stars) and SO/HET protons (cyan filled circles) and electrons (cyan filled stars). The lines depict the obtained linear fits, while the dotted lines and the corresponding shaded area with the same colouring represent the 1$\sigma$ error per fit. The peak of the SXR flare is presented as a blue triangle, the peak of the HXR flux as a magenta right triangle, the type II burst at m-$\lambda$ with a reverse orange triangle, while the DH type II burst is represented with a black X symbol. All solar eruptive signatures have been shifted; see text for details} 
\label{fig:VDA}
\end{figure}

The relation between the solar release time (SRT) $t_{\mathrm{rel}}$, the travel time $t_t$ and the onset time $t_0$ of a given particle at a given energy is:
\begin{equation}
t_0 = t_{\mathrm{rel}} + t_t, \ ~ \mathrm{with} \ ~ t_t = \frac{L}{v},
\label{eq:TSA}
\end{equation}
with $v$ the particle's speed and $L$ the apparent path length (APL).
For a spacecraft located at 1 AU, a Parker spiral modelled using a solar wind (SW) speed of 400 km/s gives an APL of $L=1.16$ AU along the magnetic field.
Using a time-shift analysis (TSA), i.e assuming a ballistic propagation of SEPs through this simple Parker spiral, the travel time $t_t$ is 47.5 min for 20 MeV protons.
With the measured onset time $t_0$ of 18:43:07 at Earth for this energy, Eq. \ref{eq:TSA} give us a SRT of 17:55:30.
By taking into account the light travel time for remote-sensing observations, this is 8 minutes after the SXR flare start and 2 minutes after the shock connection to Earth, which provides a very good initial estimate.
However, in complex events such as the present one, TSA can be not accurate enough, because the real path length $L$ is longer \citep{Vainio_2013}, or because SEPs are injected for longer durations (by a CME-driven shock for example) \citep{Reames_2021}.
To determine a more realistic SEP release times at each spacecraft, we apply a velocity dispersion analysis \citep[VDA;][and references within]{Kolympiris_2023}.
The technique relies on the assumption that the earliest detected particles across different energies are released simultaneously at the Sun and propagated to the observer without significant scattering, following a common interplanetary path length $L$ \citep[see also][for a discussion of these assumptions]{Lintunen_2004}.
A linear fit to the onset times measured at multiple energies then yields estimates of both the SEP release time at the Sun and the corresponding interplanetary path length.

\begin{figure*}[t!]
\centering
\includegraphics[width=0.66\columnwidth]{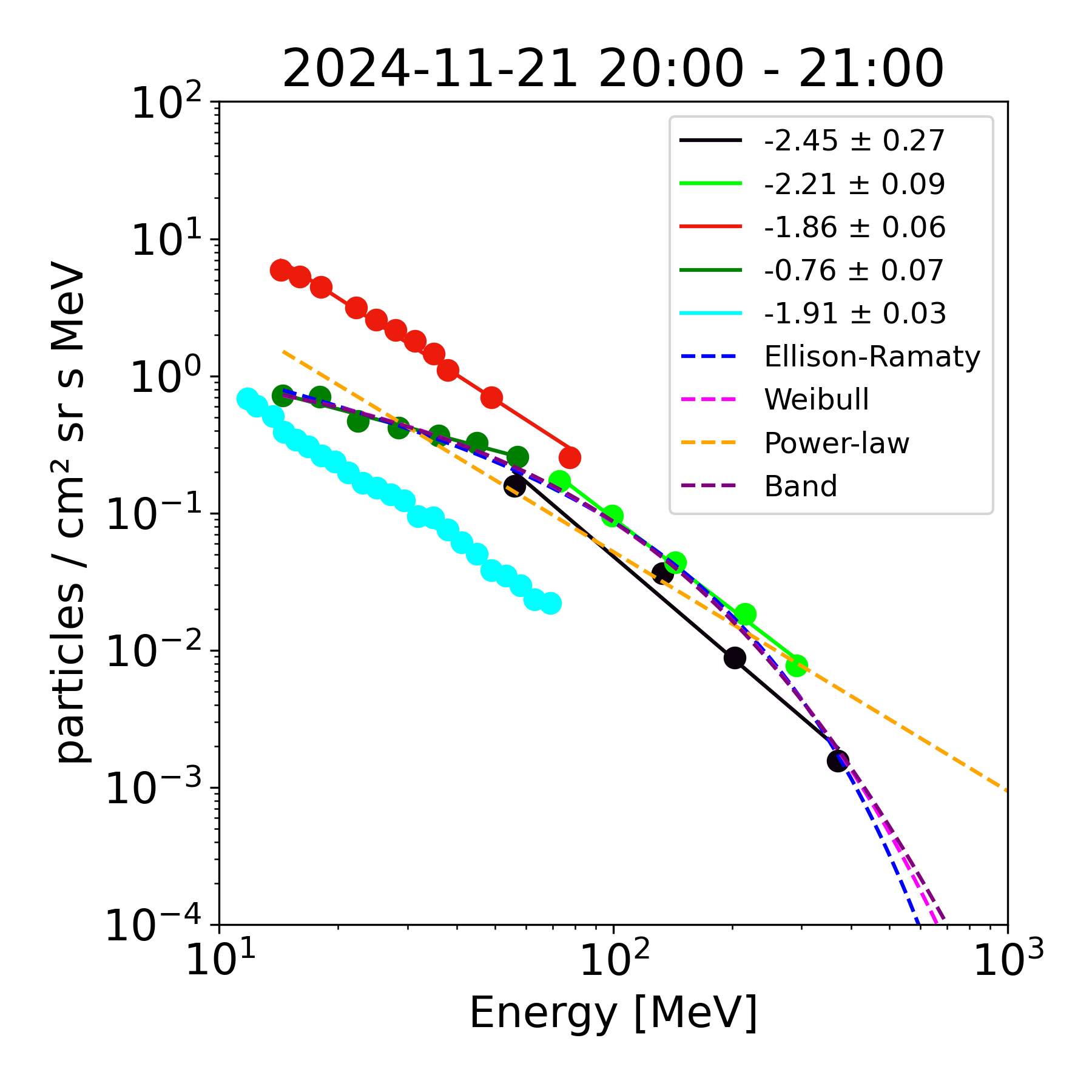}
\includegraphics[width=0.66\columnwidth]{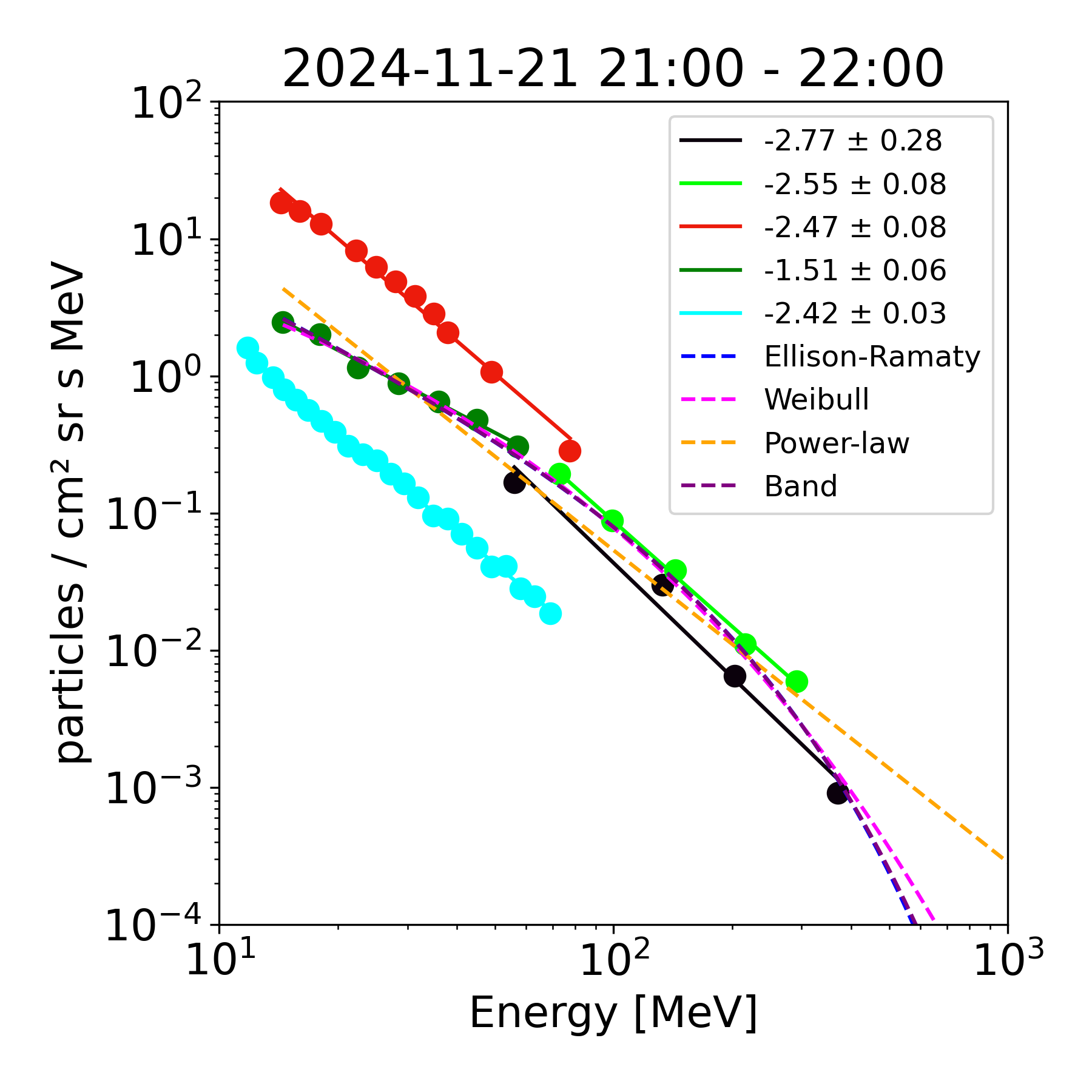}
\includegraphics[width=0.66\columnwidth]{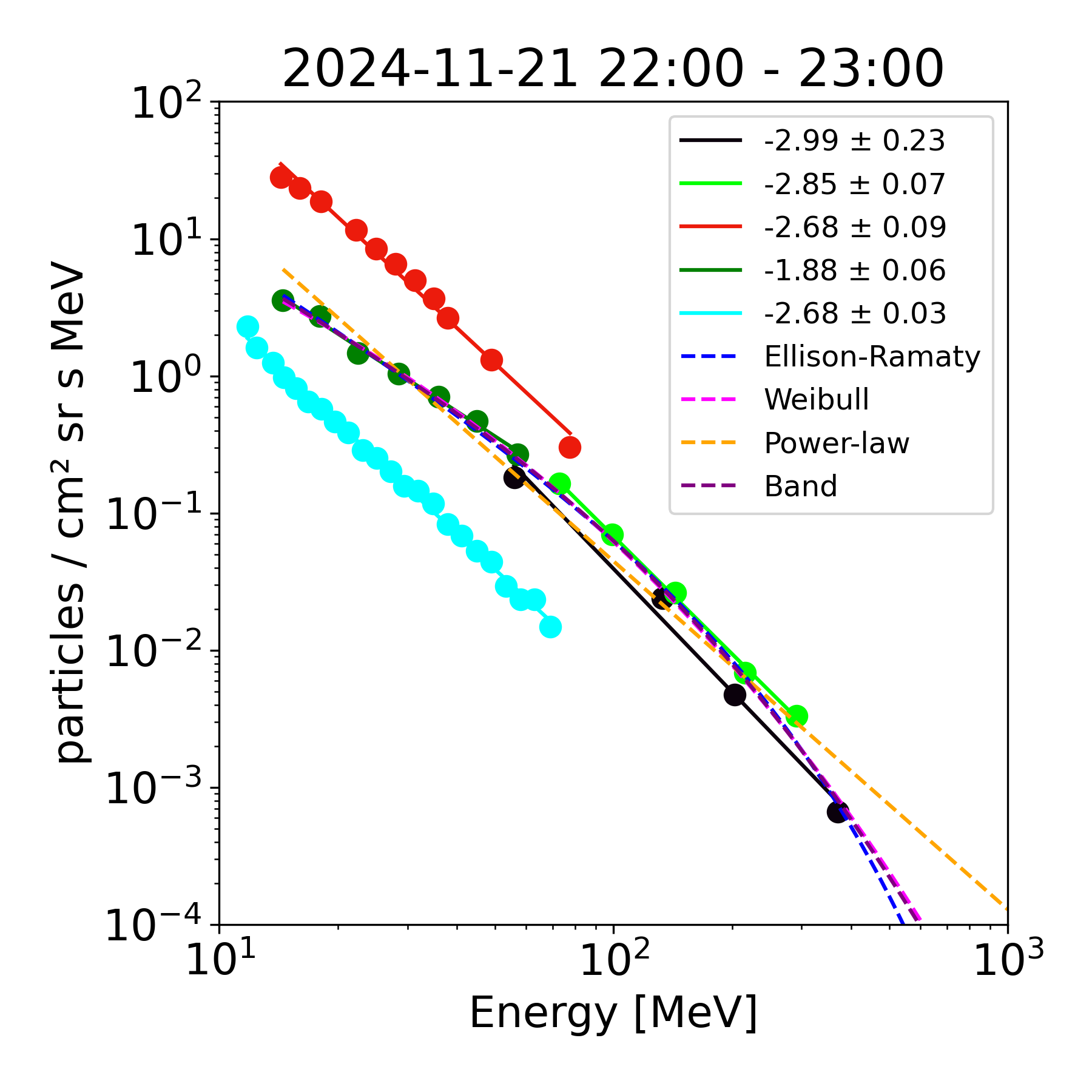}
\caption{Hourly evolution of proton energy spectra measured by SO/HET (cyan), GOES/SEISS (black), STEREO/HET (red), SOHO/ERNE (green) and SOHO/EPHIN (lime) during the 21 November 2024 SEP event, from 20:00 to 23:00 UT. Each panel displays particle flux as a function of energy (MeV) on logarithmic axes, with power-law fits overlaid for each spacecraft, separately. Additionally, the combined observations from SOHO/ERNE and GOES/SEISS were fitted using power-law, Weibull, Band, and Ellison–Ramaty models. See text for details.} 
\label{fig:spectra}
\end{figure*}

We performed VDA to the SEP measurements of STA/HET, SOHO/ERNE, and SO/HET (see Fig.~\ref{fig:VDA}).
SO high-energy measurements ($\geq$10 MeV) were used because the lower-energy measurements appeared to behave differently from them.
For STA/HET we used data covering 10-100 MeV for protons and 0.7-4 for electrons, for SOHO/ERNE we used data from 10-80 MeV for protons and SOHO/EPHIN data from 0.25-10 MeV for electrons.
For SO we used protons from 10-75 MeV and for electrons 0.7-4 MeV.
We find an SEP SRT for STA of 17:57 $\pm$3 min, for SOHO 18:11$\pm$3 min and for SO/HET 17:58 $\pm$1 min.
The VDA-derived SRTs represent the time that particles were injected at the Sun, making them directly comparable across all observing spacecraft.
The inferred SEP SRTs are similar at STA and SO, but differ from SOHO by about 15 and 13 min, respectively.
To relate these SRTs to signatures of flares and/or CMEs detected at each spacecraft, one must either convert them to the equivalent release times at the spacecraft by adding the light-travel time from the Sun to the observer or to shift back the relevant solar signatures. 
For the APLs we find $L$=1.41 $\pm$0.09 AU for STA, $L$=0.88 $\pm$0.11 AU for SOHO and $L$=1.56 $\pm$0.04 AU for SO.

\subsection{SEP energy spectra}
\label{subsect:sep_spectra}

For this high-energy multi-spacecraft SEP event, direct observations of the peak proton flux were obtained over a broad energy range from 14.4 to 500 MeV. Figure~\ref{fig:spectra} shows the differential peak proton spectrum using measurements from HET on STEREO (red filled circles), ERNE on SOHO (green filled circles), EPHIN on SOHO (lime filled circles) and SEISS on GOES (black filled circles).
The peak intensities (in units of particles/cm$^{2}$ sr s MeV, 5-min averages; 10-min averages for SOHO/EPHIN) in the prompt component were identified as the maximum values observed shortly after the in-situ event onset, excluding any energetic storm particle contribution.
For some events, the maximum intensity in the prompt component appears as a plateau in the time–intensity profile; in such cases, the peak intensity is taken as the highest value within that plateau \citep[see details in][]{Kouloumvakos_2024}.

The energy spectra measured by STA/HET over an 13.6–100 MeV and SO/HET from 11.35-75.11 MeV at 20:00-21:00, follow an inverse power-law with $\gamma$=1.86$\pm$0.06 and $\gamma$=1.91$\pm$0.03, respectively. In addition, observations from SOHO/ERNE 13-60 MeV energy range yield a spectral index of $\gamma$=0.76$\pm$0.07. At higher energies, SOHO/EPHIN measurements spanning from $\sim$76-296 MeV are also well represented by an inverse-power law yielding $\gamma$=2.21$\pm$0.09, while GOES/SEISS data from 40-500 MeV were fitted with an inverse-power law leading to $\gamma$=2.45$\pm$0.27. All energy spectra up until 100 MeV are depicted as solid lines, using the corresponding colour scheme for each observer. Combining SOHO/ERNE, SOHO/EPHIN and GOES/SEISS ion measurements, four fits were applied: (a) an inverse power-law (dashed orange line), (b) an Ellison--Ramaty \citep[dashed blue line;][]{Ellison_1985}, (c) a Weibull \citep[dashed green line;][]{Laurenza_2015} and (d) a Band fit \citep[dashed purple line;][]{Band_1993}.

As the event progresses into the 21:00–22:00 UT interval, a clear steepening of the spectra is observed across all instruments. The STA/HET and SO/HET measurements indicate higher spectral indices ($\gamma$=2.47$\pm$0.08 and $\gamma$=2.42$\pm$0.03, respectively), reflecting a softer energy distribution, while SOHO/ERNE likewise shows a marked increase in $\gamma$ (i.e. 1.51$\pm$0.06). Concurrently, the flux levels increase over the entire energy range, and the high-energy tail extends to higher energies, suggesting continued particle acceleration and improved magnetic connectivity. The high energy measurements from SOHO/EPHIN and GOES/SEISS remain consistent with this trend, exhibiting progressively steeper spectra with $\gamma$=2.55$\pm$0.08 and $\gamma$=2.77$\pm$0.28, respectively.

By 22:00–23:00 UT, the spectra reach their highest intensities and display further softening, with the largest spectral indices recorded during this interval (STA/HET: 2.68$\pm$0.09; SO/HET: 2.68$\pm$0.03; SOHO/ERNE: 1.88$\pm$0.06; SOHO/EPHIN: 2.85$\pm$0.07; GOES/SEISS: 2.99$\pm$0.03). The enhancement is particularly pronounced at higher energies, where the rollover becomes more evident and shifts toward higher energies. The spectral evolution therefore suggests continued particle acceleration during the event, but its interpretation is likely observer dependent. STA was comparatively well connected to the flare site and to the early CME-driven shock, while SO appears to sample a population with spectral evolution broadly similar to STA. In contrast, the near-Earth measurements include a higher-energy component observed under a more uncertain magnetic connection. Thus, rather than implying a single uniform acceleration history across all instruments, the spectra point to a combined effect of ongoing acceleration and observer-dependent transport/connectivity. \citep[e.g.][]{Reames_1999, Desai_2016}. While simple power-law fits reproduce the low- to intermediate-energy range, they systematically overestimate high-energy fluxes. In contrast, curved models such as the Ellison–Ramaty, Weibull, and Band functions provide a more accurate representation, reflecting finite acceleration timescales and energy-dependent escape processes \citep{Ellison_1985, Band_1993, Martucci_2023}. The resulting reduced $\chi^{2}$ is 0.72 for the Ellison-Ramaty, 0.75 for the Band, 0.71 for the Weibull and 2.12 for the power-law fits.
Clearly, the power-law model fit fails to capture the evolution of the data, particularly at high energies. In contrast, the other fits show good overall agreement, with the Weibull fit appearing to be the most optimal, as it yields the lowest reduced $\chi^{2}$. The coexistence of spectral softening and an increasing cutoff energy points to a complex interplay between acceleration and transport, where continued energization at the shock is modulated by interplanetary propagation effects, including scattering and evolving magnetic connectivity \citep{Ng_2008,Vainio_2014}.
Overall, the observed spectral evolution for near-Earth is consistent with sustained shock acceleration shaped by transport processes, in agreement with recent studies of relativistic particles \citep{Martucci_2023, Kouloumvakos_2024}.
Although the near-Earth magnetic connection is more uncertain (see Sect. \ref{subsect:MHD_parameters}), the combined spectral behaviour remains qualitatively compatible with CME-driven shock acceleration, while the similarities between STA and SO suggest that their observed spectra may be shaped by comparable connectivity and transport conditions.

\section{3D reconstruction of the coronal shock and magnetic field lines}
\label{sect:3D_reconstrcution}

Deriving spacecraft magnetic field lines and their footpoints on the Sun surface allows to estimate the source position of the SW and energetic particles measured by these different spacecraft.
In our study, to investigate the conditions under which particle acceleration may have occurred, we also reconstruct the three-dimensional geometry of the coronal shock wave.
Magnetic connectivity between the shock and the different observing spacecrafts was then evaluated by tracing field lines outwards from the solar surface.

In this section, we first describe the different approaches used to derive the magnetic field lines connected to the spacecraft. We then present the reconstruction of the expanding coronal shock from remote sensing observations. Finally, by combining the reconstructed shock geometry with a global magnetohydrodynamic (MHD) simulation of the corona, we derive the evolution of the shock parameters along the magnetic field lines connected to the observers.

\subsection{Magnetic field line reconstruction}
\label{subsect:mag_connectivity}

\begin{table*}[t]
\caption{Location of observing spacecraft, measured solar wind (SW) speed, connection angle to SXR flare, and derived magnetic footpoints with the three presented methods along with their angular distances from the SXR flare.}
    \centering
    \small
    \begin{tabular}{l|ccc|c|c|ccc|ccc|ccc}
     & \multicolumn{3}{c|}{Spacecraft location} & SW & CA &\multicolumn{3}{c|}{Parker spiral footpoint} & \multicolumn{3}{c}{PFSS footpoint} & \multicolumn{3}{c}{MHD footpoint} \\
    Name & CRLN & CRLT & R & speed & $\Delta \phi$ & CRLN & CRLT & $\Delta \alpha$ & CRLN & CRLT & $\Delta \alpha$ & CRLN & CRLT & $\Delta \alpha$\\
    \hline
    STA & 274.9 & -1.6 & 0.97 & 300 & 32.1 & 333.9 & -1.6 & 14.8 & 341.1 & 4.6 & 27.8 & 352.6 & -7.7 & 11.6\\
    Earth & 246.9 & 2.0 & 0.99 & 410 & 58.8 & 307.2 & 2.0 & 59.2 & 285.8 & -10.8 & 79.0 & 281.5 & -11.8 & 19.1\\ 
    SO & 242.7 & 6.3 & 0.83 & 420 & 72.8 & 293.2 & 6.3 & 75.5 & 283.8 & -11.7 & 80.8 & 274.6 & -10.0 & 90.5\\
    PSP & 158.7 & 3.8 & 0.7 & 400* & 164.8 & 201.2 & 3.8 & 164.3 & 243.7 & -14.0 & 118.6 & 223.8 & -8.1 & 139.2\\
    Bepi & 181.4 & 3.6 & 0.36 & 400* & 162.9 & 203.1 & 3.6 & 162.4 & 243.7 & -14.1 & 118.7 & 224.5 & -8.2 & 138.5\\
\end{tabular}
\tablefoot{Heliographic Carrington longitude (CRLN) and latitude (CRLT) in degrees, radial distance (R) in AU, SW speed in km/s.
The connection angle to flare is $\Delta \phi = \phi_{flare}-\phi_{\mathrm{FP}}$ in degrees.
The corresponding magnetic footpoints (FP) are derived assuming, respectively, a Parker spiral configuration, a PFSS model, and a MHD cube.
The angular distance to flare is calculated using: $\cos\Delta \alpha=\sin(\mathrm{lat}_{flare})\times\sin(\mathrm{lat}_{\mathrm{FP}})+\cos(\mathrm{lat}_{flare})\times\cos(\mathrm{lat}_{\mathrm{FP}})\times\cos(\mathrm{lon}_{flare}-\mathrm{lon}_{\mathrm{FP}})$ in degrees.}
\label{tab:fps_and_flare_locations}
\end{table*}

We made use of three different methods to reconstruct spacecrafts magnetic field lines.
The Parker spiral model is the simplest and easiest to compute.
It provides a simple analytical description of the interplanetary magnetic field, assuming a steady, radially expanding SW and a rotating Sun.
However, it does not account for the complex magnetic field structure in the corona or for temporal and spatial variations in the SW.
This is the reason why we also used two other magnetic field models.

\begin{figure}[t]
\centering
\includegraphics[width=\columnwidth]{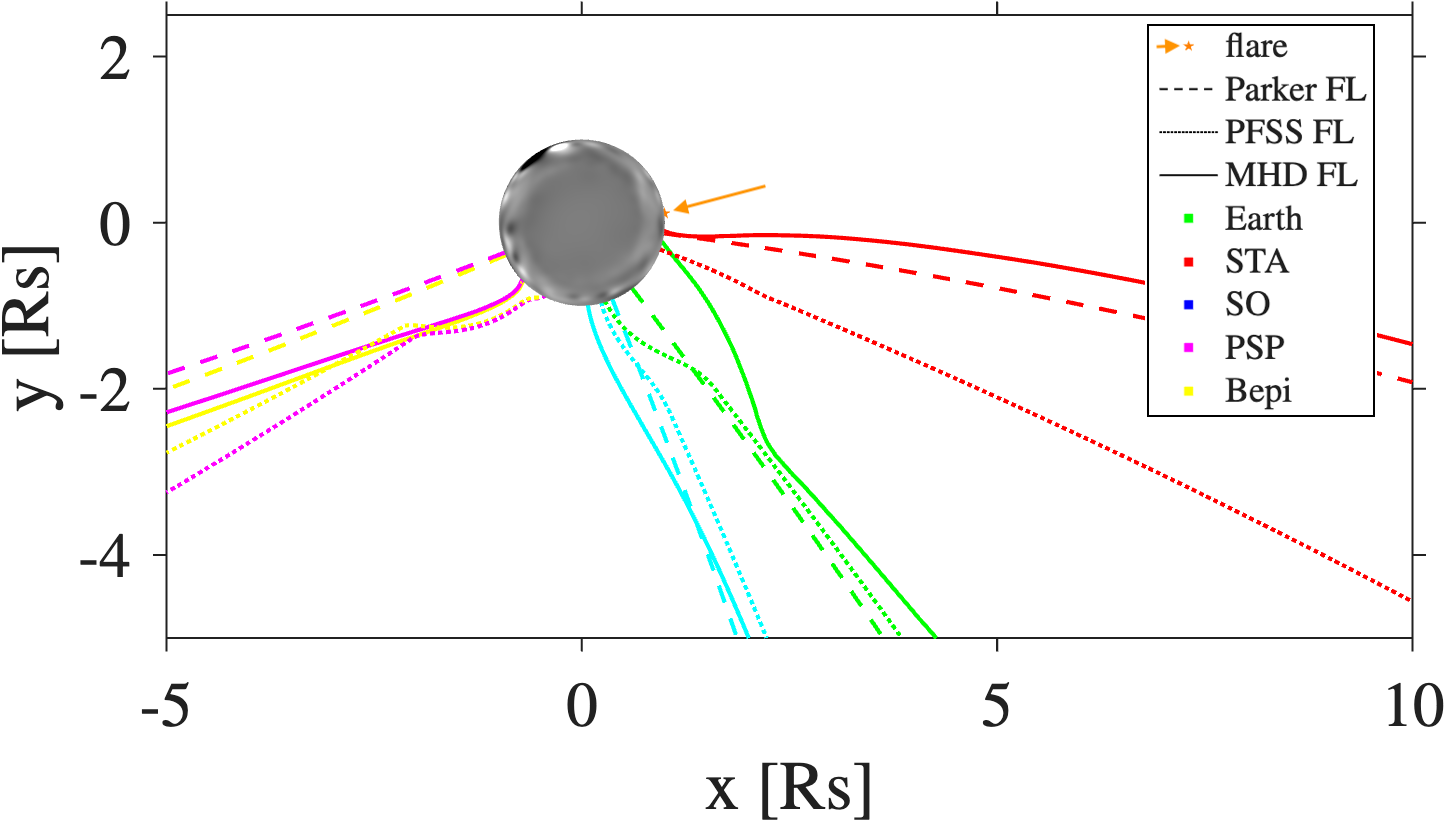}
\caption{Representation of the spacecraft magnetic field lines in 3D viewed from the solar north. Dashed, dotted, and plain lines represent, respectively, Parker spiral, PFSS and MHD field lines, coloured differently according to the spacecraft (green: Earth, red: STA, blue: SO, purple: PSP, yellow: Bepi). The orange star pointed by the arrow represent the flare location, and the Sun is coloured according to the magnetic field intensity from the ADAPT magnetogram model.} 
\label{fig:3D_FLs}
\end{figure}

\begin{figure}[t]
\centering
\includegraphics[width=\columnwidth]{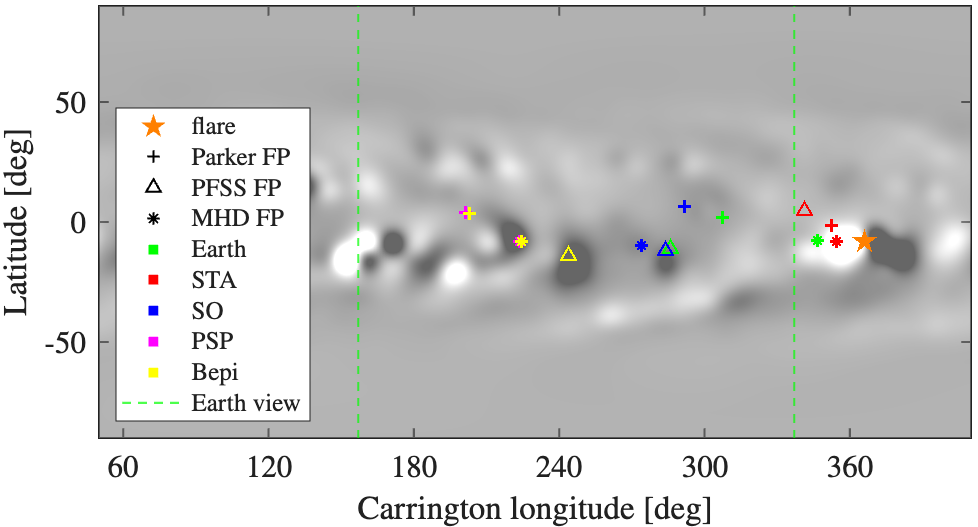}
\caption{Magnetogram of the surface of the Sun at the time of the event. The colour from white to black reflect the magnetic field intensity at the surface of the Sun. The crosses, triangles, and asterisks represents respectively the Parker spiral, PFSS, and MHD footpoints, of the Earth (in green), STA (in red), Solar Orbiter (SO, in blue), PSP (in magenta) and Bepi (in yellow). The orange star indicates the location of the EUV flare, and the green dotted lines frame the part of the Sun’s surface visible from Earth.} 
\label{fig:magnetogram}
\end{figure}

The potential field source surface (PFSS) model is a magnetostatic reconstruction technique of the magnetic field between 1 and 2.5 $R_{\odot}$, using a photospheric magnetogram as its lower boundary, and assumes that the corona is in current-free equilibrium state \citep{Wiegelmann_2012}.
The PFSS model is one of the simplest approaches to model the magnetic field of the solar corona.
The approximations are therefore: $\nabla \wedge \textbf{B} = 0$ everywhere, the magnetic field is radial at $R = 1$ Rsun, i.e on the Sun surface, and at the surface source, $R = R_s$, typically $R = 2.5$R$_\odot$.
The PFSS coronal model is usually supplemented by an interplanetary model, in our case a Parker spiral \citep{Owens_2013}.
We used the Magnetic Connectivity Tool (\url{http://connect-tool.irap.omp.eu/}) developed by IRAP \citep{Rouillard_2020, Poirier_2021}, a free online tool, to derived 100 magnetic field lines using the PFSS model and Air Force Data Assimilative Photospheric flux Transport (ADAPT) magnetogram model.

In the last approach, a nominal Parker spiral is computed from each spacecraft position to a radial distance of $20\,R_{\odot}$.
Below this altitude, the magnetic field topology was derived using a three-dimensional magneto-hydrodynamic (MHD) model of the background solar corona, the magneto-hydrodynamic around a sphere thermodynamic (MAST) cubes developed by Predictive Science Inc \citep[PSI;][]{Lionello_2009}, driven at its inner boundary by an HMI/SDO synoptic magnetogram representative of the period of the event.
Figure~\ref{fig:3D_FLs} shows simultaneously the three different type of magnetic field lines derived to assess magnetic connectivity of spacecrafts, i.e Parker spiral (dashed lines), PFSS+Parker spiral (dotted lines) and MHD+Parker spiral (plain lines).

Table \ref{tab:fps_and_flare_locations} lists the main magnetic footpoints obtained using the Parker spiral, PFSS, and MHD models. 
The Parker spiral preserves the spacecraft latitude, resulting in positive latitudes for all observers except STA, whereas the PFSS and MHD models yield negative latitudes.
STA is an exception in the PFSS model, with a positive latitude, however, this result is model-dependent, as Solar-MACH \citep{Gieseler_2023} yields a negative value. 
Figure~\ref{fig:magnetogram} shows the corresponding footpoint locations.
The associated field lines provide a first-order estimate of the nominal magnetic connection between each observer and the source region.

\subsection{Coronal shock wave modelling}
\label{subsect:shock_modelling}

Fast CMEs can generate pressure fronts capable of steepening into shock waves in the low corona.
The three-dimensional reconstruction of such pressure fronts enables the tracking of their expansion beyond projection effects and provides estimates of their geometry and kinematics in the corona.
In this study, we apply the ellipsoidal fitting technique originally introduced for multipoint coronal imaging \citep[e.g.][]{Kwon_2015, Rouillard_2016, Kwon_2017} to model the evolution of the pressure front associated with the 21 November 2024 eruption.
The method consists in identifying the outermost perturbation region produced by the CME in both EUV and white-light running-difference imagery.
This region, typically characterised by an expanding bubble structure off-limb and a coronal wave on-disk, is assumed to represent the boundary of the pressure wave surrounding the CME.
A triaxial ellipsoid is then fitted to this boundary, parametrised by its three half-axes: $a(t)$ along the radial direction, and $b(t)$ and $c(t)$ describing the lateral expansion.
In this work we make use of remote sensing EUV, coronographic and heliospheric observations of the solar corona obtained from STA/EUVI, COR1, and COR2, SOHO/LASCO/C2 and C3.

In STA/EUVI 171$\AA$ at 17:55, a bright coronal loop is visible and start to expand, forming a shock at 17:57:30.
The event was not visible from SDO/AIA.
The CME-driven shock appears in STA/COR1 at 18:01, STA/COR2 at 18:23, SOHO/LASCO/C2 at 18:12, and SOHO/LASCO/C3 at 18:30.
It was possible to fit an ellipsoid to the shock wave front from 17:57:30 to 18:53:30, i.e when the shock front reached $\sim$10.5 AU.
Due to the observational configuration during the event, the reconstruction is subject to limitations.
STA and near-Earth observers were separated by a small longitudinal angle, providing nearly redundant viewpoints.
SO, although well positioned longitudinally, does not host a coronagraph, and the Wide-field Imager for Solar Probe camera \citep[WISPR;][]{Vourlidas_2016} onboard PSP did not detect the CME in this case.
As a result, no clear view of the western flank of the CME was available, reducing the geometric constraints on the ellipsoid and increasing the uncertainty in the fitted lateral expansion.
Thus, additional constraints had to be used to achieve a 3D fit of the shock. Based on the statistical results obtained by \citet{Jarry_2023}, we assumed a radial direction of propagation from the projected centre of the shock, with a perfectly spherical shock front extension.
The centre of the reconstructed shock was derived from the flare location.
In Carrington coordinates, the longitude was fixed to $6^{\circ}$ and the latitude evolved from $-11^{\circ}$ to $8^{\circ}$.
The resulting reconstructed shock surface is shown in Fig.~\ref{fig:shock_fit} of the Appendix \ref{annex: additional_material}, represented by an orange ellipsoid overlaid on STA EUV and white-light images.
The shock front of this modelled shock wave is propagating at a maximum 3D speed of 2518 km/s, while extending laterally at a speed of 1730 km/s.
Their temporal evolution, such as that of the other shock geometric parameters (height of the shock front, lateral extension), is consistent with the results of \citet{Jarry_2023}.

\subsection{Derivation of shock MHD parameters}
\label{subsect:MHD_parameters}

\begin{figure}[t]
\centering
\includegraphics[width=\columnwidth]{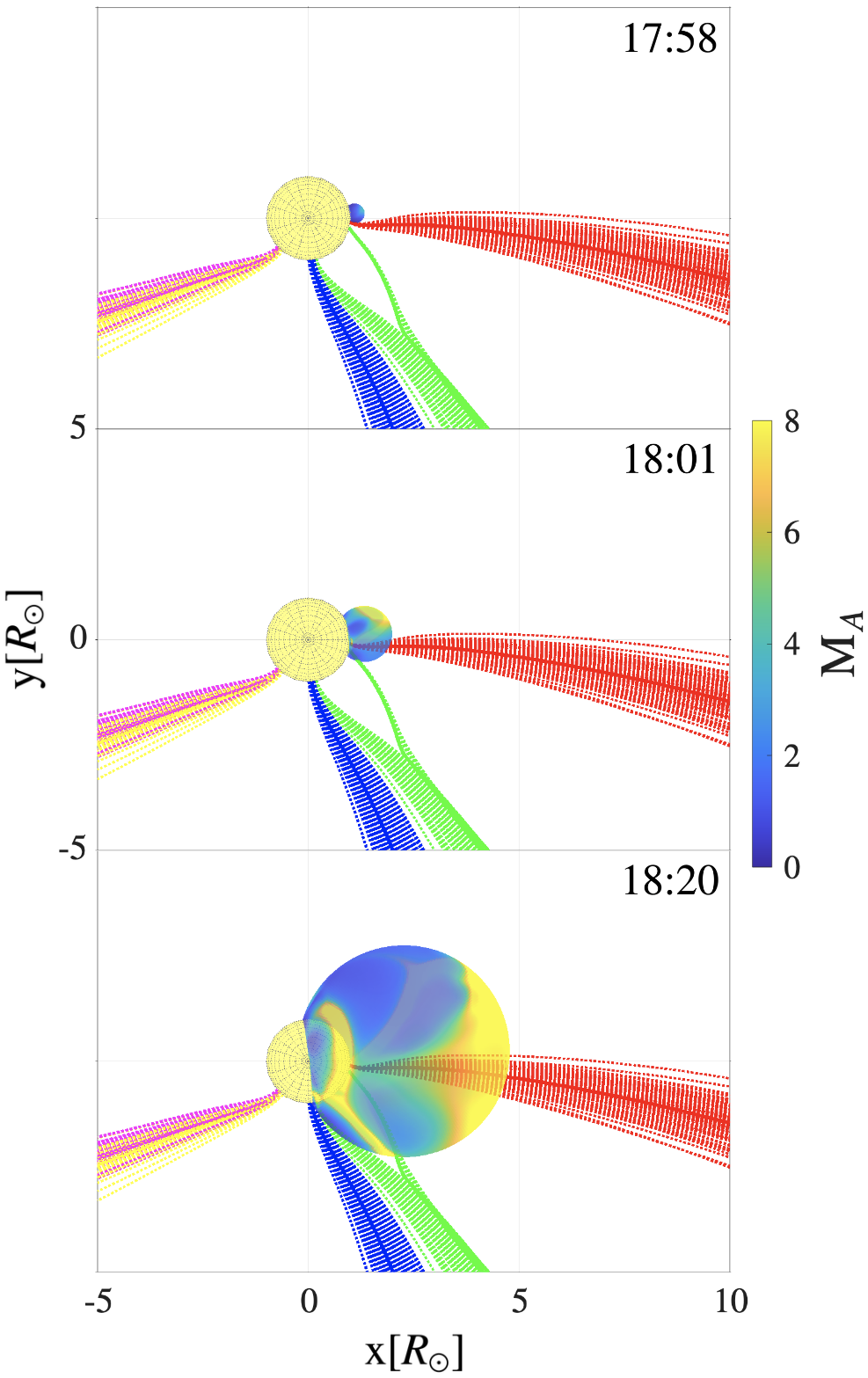}
\caption{View of the ecliptic plane as seen from north, showing the 3D shock model and the magnetic connectivity. For each panel the yellow sphere represents the Sun in true scale, the coloured ellipsoid represents the reconstructed shock wave, coloured according to the intensity of the Alfvénic Mach number M$_A$. The lines represent the magnetic field lines connected to Earth (green), STA (red), SO (blue), PSP (pink), and Bepi (yellow, behind PSP). For each spacecraft, hundred connectivity lines are created according to a random number of latitude and longitude taken in a normal distribution around main latitude and longitude at 25 solar radius, start of the MHD cube. From top to bottom, the panels show the shock position at 17:58, i.e. at the connection to STA field lines; at 18:01, i.e. at the connection to the first packet of Earth field lines, and at 18:20, i.e. at the connection to SO field lines (times in UTC).}  
\label{fig:3d_shock}
\end{figure}

To estimate the properties of the pressure front, the reconstructed ellipsoid was combined with the MHD cube described previously (see Sect. \ref{subsect:mag_connectivity}).
This model takes into account the detailed thermodynamics with realistic energy equations accounting for thermal conduction parallel to the magnetic field radiative losses, and coronal heating, as the effect of the Alfvén waves on the expanding coronal plasma.
Thanks to this comprehensive energy calculations, realistic SW mass fluxes can be attained, and this allows to derive the distribution of the coronal density and temperature.
This enables the determination of plasma parameters at each point on the reconstructed surface, including magnetic field vectors and characteristic wave speeds.
Figure \ref{fig:3d_shock} shows the modelled CME-shock surface coloured according to the Mach Alvénic number intensity for three different timings: 17:58, 18:01, and 18:20 UTC.
To account for uncertainties in the SW speed and in the coronal magnetic configuration, an ensemble of field lines was generated for each spacecraft.
Following a Gaussian distribution with a width of approximately $8^{\circ}$ in both latitude and longitude at $20\,R_{\odot}$, 99 additional field lines (coloured dashed lines in the figure) were produced around the nominal one (plain line).
This angular spread corresponds to variations in the SW speed of $\sim 25 \pm 5\,\mathrm{km\,s^{-1}}$ at 1~AU.
The shock properties were then evaluated at the intersections between the reconstructed surface and each member of the field-line ensembles derived previously.
This combined modelling allows the extraction of key shock parameters along magnetically relevant regions of the pressure front, including the shock normal angle $\theta_{\mathrm{Bn}}$, the Alfvénic Mach number M$_\mathrm{A}$ and the local shock speed $V_{\mathrm{sh}}$.

\begin{table*}[t]
\caption{Timings at which the reconstructed 3D shock connects with the magnetic field lines associated with different spacecraft, and values of the shock parameters at these intersection points within their interval of values.}
    \centering
    \small
    \renewcommand{\arraystretch}{1.2}
    \begin{tabular}{l|cc|ccccc}
    & \multicolumn{2}{c|}{Connection} & \multicolumn{5}{c}{Shock parameters at connection time} \\
    Name & Time & Interval & R$_{sh}$ [AU] & V$_{sh}$ [km/s] & $\theta_{Bn}$ [deg] & M$_A$ & X$_R$\\
    \hline
    STA & 17:58 & 17:57-17:59 & 1.15 [1.06--1.23] & 691 [577--805] & 76.7 [69.8--83.6] & 0.09 [0.04--0.13] & 0.01 [0.0--0.03]\\
    Earth & 18:01 & 17:59-18:20 & 1.06 [1.00--1.08] & 587 [554--620] & 77.5 [76.0--79.0] & 0.11 [0.10--0.13] & 0.02 [0.01--0.03]\\
    SO & 18:20 & 18:18-18:23 & 1.03 [1.00--1.23] & 547 [528--565] & 74.7 [68.1--81.2] & 0.17 [-0.38--0.73] & 0.05 [-0.60--0.70]\\
\end{tabular}
\label{tab: FL_connection}
\end{table*}

\begin{figure}[t]
\centering
\includegraphics[width=\columnwidth]{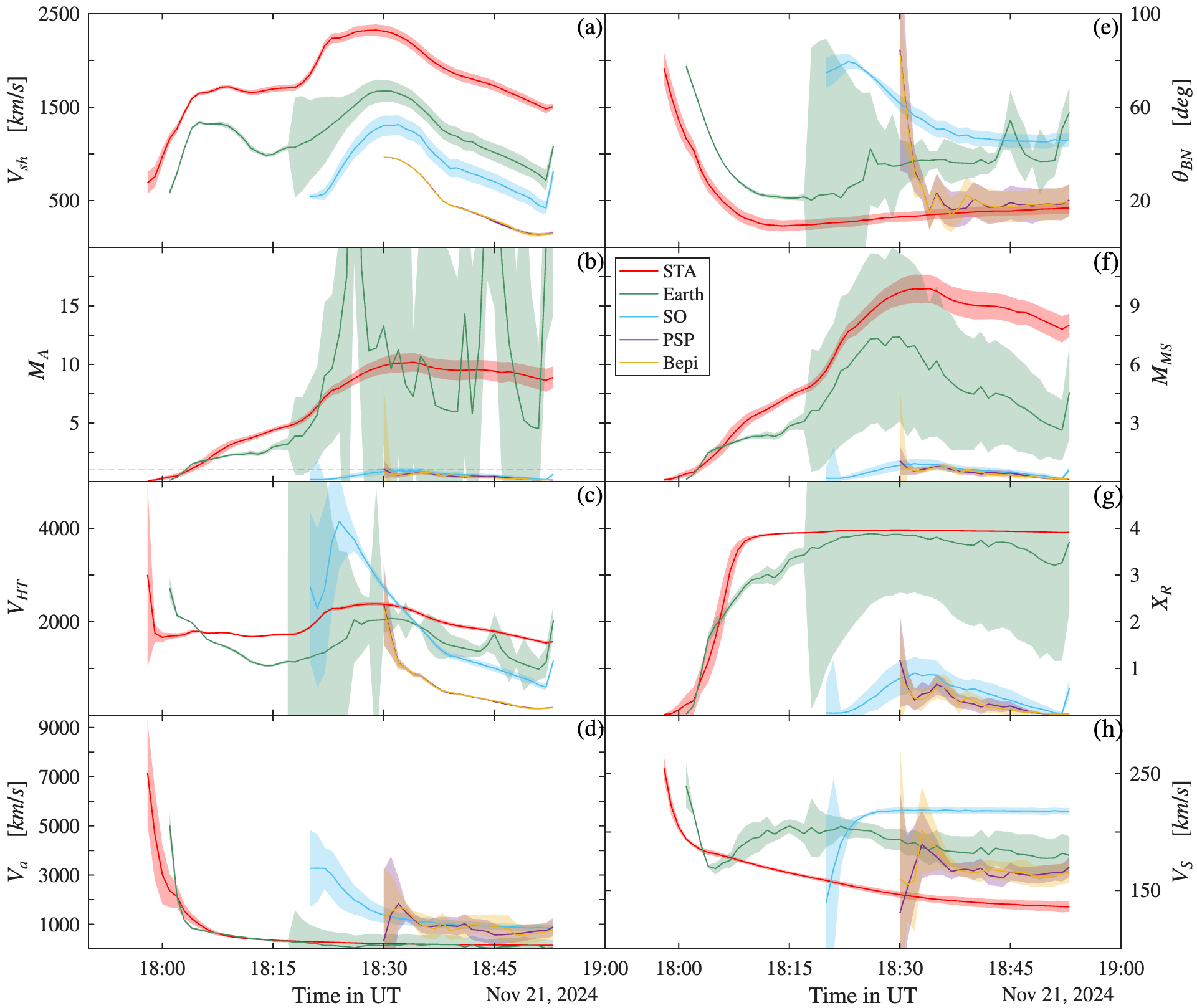}
\caption{Evolution of shock parameters along the different magnetic field lines. Panel (a): Shock speed [km/s]. Panel (b): Alfvénic Mach number (M$_\mathrm{A}$). Panel (c): Hoffmann-Teller speed [km/s]. Panel (d): Alfvénic speed [km/s]. Panel (e): $\theta_{BN}$ [deg]. Panel (f): Magnetosonic Mach number (MMS). Panel (g): shock compression ratio. Panel (h): Sound speed [km/s]. Uncertainties are derived using the standard deviation of a sample of 100 magnetic field lines (shown in Fig.~\ref{fig:3d_shock}, see the text for details) and the plain line represent the main MHD field line for each spacecraft, shown in Fig.~\ref{fig:3D_FLs}. }  
\label{fig:shock_params}
\end{figure}

Figure \ref{fig:shock_params} shows the evolution of shock parameters with time along the different magnetic field lines connected to each spacecraft.
The shock connects early with STA magnetic field line at 17:58, and then with Earth magnetic field line at 18:01.
SO follows at 18:20, and PSP and Bepi connects later, both at 18:30.
The uncertainties associated with these connection timings, derived using the hundred lines, as well as the shock parameters values at the intersection between the shock and the magnetic filed lines at the connection time, are described in Table \ref{tab: FL_connection}.
The wide variation in the connection timings between the shock and the Earth's field line is due to the division of the 100 field lines into two packets (see the green lines on Fig. \ref{fig:3d_shock}).
The first group is connected to a source very close to the flare, the same active region to which STA is connected (Fig. \ref{fig:magnetogram}), while the second group has the same magnetic connection as SO.
M$_A$ become super-Alfvénic ($>1$) along Earth and STA field lines at the same time, i.e 18:04, whereas it remains subcritical for the others.
Shock parameters evolution for Earth and STA follow each other closely and presents similar intensities, whereas these ones are fainter on PSP, Bepi and SO.
We note that the M$_A$ evolution along Earth field line is very perturbed (green line on panel (b) of Fig.~\ref{fig:shock_params}), and probably reveal a magnetic connection close to the heliospheric current sheet (HCS), consistent with the division of field lines in two groups.
The shock parameters derived along the STA-connected field line from the 3D reconstruction are considerably less affected by fluctuations.
The Magnetosonic Mach number $M_{MS}$ is stable for both of them and reach the maximum value of 9.9$\pm$0.7 along STA field line and 7.4$\pm$4.3 along Earth one, at 18:35 and 18:31, respectively.
The magnetic field is quasi-perpendicular for both spacecraft until $\sim$ 18:15 UT and then $\theta_{\mathrm{Bn}}$ remains under 20$^\circ$ for STA field line and oscillate between 20 and 60 for Earth field line.

\section{SEPs acceleration analysis}
\label{sect:acceleration_analysis}

To investigate the respective contributions of flare- and CME-driven shock-related acceleration to the observed SEP event, we examine the timing and magnetic connectivity of the particle release, the evolution of SEP intensities and composition, and the associated radio emissions as signatures of this acceleration.

\subsection{Event injection timing and SEP intensity evolution}
\label{subsect:injection_intensity_evolution}

STA exhibits a very good magnetic connection to the source region (See Fig. \ref{fig:magnetogram}).
It is well connected to the flare site, allowing early access to flare-accelerated particles.
As the CME expands, this connection extends to the shock front, enabling STA to efficiently sample shock-accelerated particles from an early stage.
The magnetic connection is maintained over an extended period, which likely accounts for the plateau observed in the SEP intensity profiles (Fig. \ref{fig:SEP_and_obs}).
During this phase, STA remains connected to shock regions characterized by high $M_A$ and quasi-perpendicular geometries ($\theta_{BN} \sim 90^\circ$), conditions known to favour efficient particle acceleration \citep{Kouloumvakos_2019, Dresing_2022}.
This suggests that particles detected at early times already include a substantial contribution from shock acceleration (see Sect. \ref{subsect:MHD_parameters}).

As discussed in Sect. \ref{subsect:MHD_parameters}, the near-Earth magnetic connection is likely located close to the HCS, leading to strong variations in the derived shock properties, particularly the M$_A$.
This limits the interpretation of the relationship between shock parameters and SEP spectral evolution at Earth, but does not invalidate the observed spectral evolution itself (Sect. \ref{subsect:sep_spectra}).
The reconstructed shock nevertheless reaches high $M_A$ values along the near-Earth field lines, indicating highly super-Alfvénic conditions favourable for efficient particle acceleration.
The magnetic field lines are distributed over two distinct groups (Fig. \ref{fig:3d_shock}), with a connection times to the shock of 18:01 and 18:17, respectively.
The first group connects to a region close to the flare site and to the magnetic connection of STA, whereas the second connects to another active region, also connected to Solar Orbiter.
This configuration provides Earth with access to particles accelerated at different locations, including the flare region.

In contrast, Solar Orbiter is magnetically connected to a region far from the flare site.
Despite this connection, its VDA release time is close to those inferred at STA and to the flare onset (Sect. \ref{subsect:sep_obs}).
This indicates that the particles detected at SO were released early in the event, although the reconstructed magnetic connection does not provide direct access to the flare region.
The inferred release time precedes the connection to the reconstructed shock front at SO (Sect. \ref{subsect:MHD_parameters}), limiting the contribution that can be attributed to local shock acceleration at the time of the initial particle release.
The SEP composition measured at SO also exhibits signatures consistent with flare-related acceleration, supporting the possibility of an early flare contribution.
An alternative scenario is that particles were initially accelerated in a confined flare near the SO-connected footpoint and subsequently released through magnetic reconnection triggered by the main eruption \citep[][]{Klein_2022, Jarry_2024}, which may be consistent with both the early VDA release time and the observed composition, although it remains speculative.
Figure~\ref{fig:timeline} in the Appendix \ref{annex: additional_material} allows to compare the shock connection timings with the ones of the VDAs results and the different solar associations timings.

\begin{figure}[t]
\centering
\includegraphics[width=0.99\columnwidth]{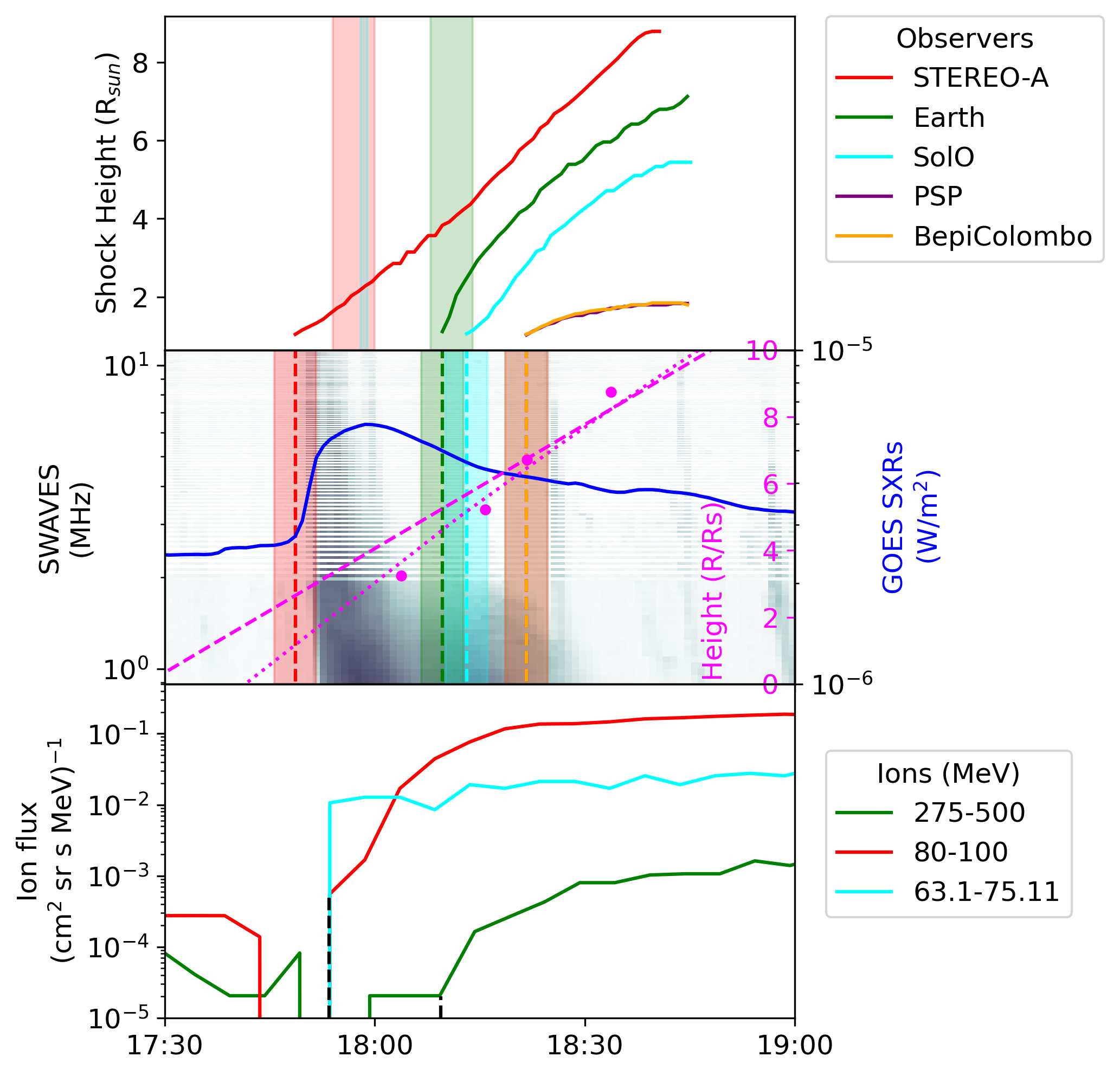}
\caption{The reconstructed shock heights by CME-modelling for STA (red), Earth (green), SO (blue), PSP (magenta) and Bepi (yellow).
The colour coded shaded areas depict the obtained results of the VDA for STA, SO and Earth (top panel).
The SXR time profile (blue line - axis on the right hand side) indicating the C-class flare, the height-time profile recovered by SOHO/LASCO (magenta line - axis on the right hand side) for the driving CME, depicted on the radio spectrum from SWAVES.
The dashed red and green lines provide the time of the intersection of the expanding shock to the connecting FL, the coloured shaded areas assume a $\pm$3min error (middle panel).
Time profiles of the high-energy SEPs measured onboard GOES (275-500 MeV; green line), STA/HET (60-100 MeV; red line) and SO/HET (63.1-75.11 MeV; cyan line) (bottom panel).
Dashed black vertical lines correspond to the onset times of the high-energy SEP measurements at each spacecraft.
All times in all panels are shifted at the Sun.}  
\label{fig:combined_plot}
\end{figure}

\begin{figure}[t]
\centering
\includegraphics[width=0.9\columnwidth]{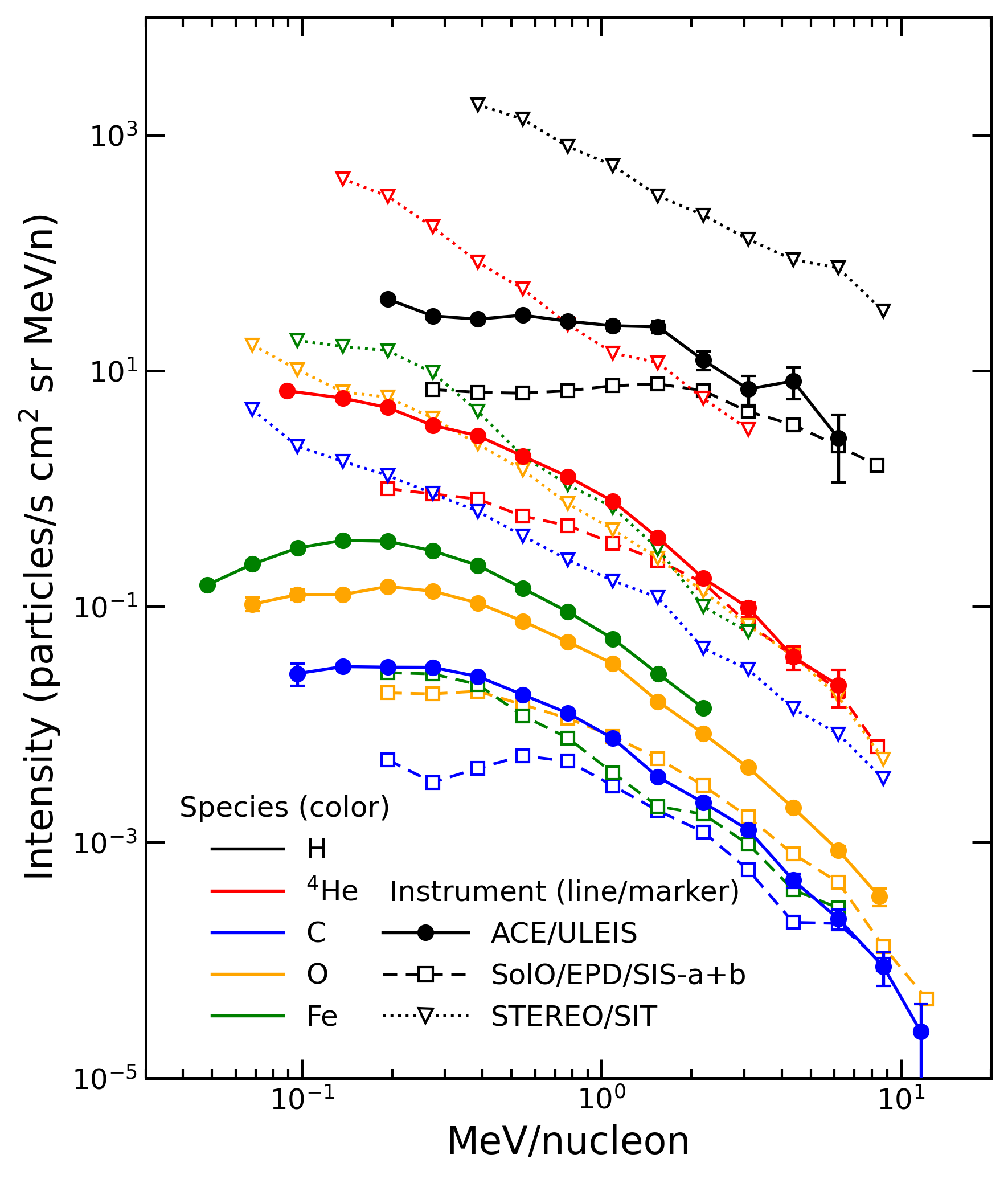}
\caption{Event integrated particle spectra from STA/SIT (dotted line + triangles), ACE/ULEIS (plain line + dots), and SO/EPD/SIS-a+b (dashed line + square).
Differential intensities of H (black), $^{4}$He (red), C (blue), O (yellow), and Fe (green) are shown as a function of energy per nucleon (MeV/nucleon).
The spectra were integrated over the periods when the 21 November event dominated the intensities, and are different at the different spacecraft (STA: 21 Nov 18:00 to 25 Nov 0:0; ACE: 21 Nov 17:00 to 24 Nov 0:0 ; SO: 21 Nov 19:00 to 23 Nov 0:0).
Some low energy points on ACE and SO are not shown since at these energies the GLE event did not produce a significant enhancement over prior activity.
The spectra provide a multi-spacecraft comparison of the
elemental composition and energy dependence of the SEP
population observed at different heliospheric locations.
}  
\label{fig:composition}
\end{figure}

Figure~\ref{fig:combined_plot} presents a multi-panel comparison between the temporal evolution of the CME-driven shock, the associated radio and flare signatures of the solar eruption and the onset of high-energy SEPs observed at Earth (GOES), STA and SO.
The figure combines remote-sensing diagnostics with particle measurements to investigate the timing and efficiency of shock acceleration during the November 21, 2024 SEP event.
In more detail, the upper panel of Fig.~\ref{fig:combined_plot} shows the reconstructed shock heights along magnetic field lines connected to several observers: STA (red), Earth (green), SO (blue), PSP (purple) and Bepi (yellow).
The shock height is expressed in solar radii (R$_{\odot}$ and shows the continued outward expansion of the CME-driven shock through the corona.
Among all connection paths, the shock propagates furthest and fastest toward STA, reaching nearly 9 $R_\odot$ by the end of the displayed interval, while the Earth-connected flank reaches approximately 7 $R_\odot$.
SO connection exhibits an intermediate evolution, whereas PSP and Bepi connections remain at comparative lower heights, not exceeding about 2 $R_\odot$.
The colour-coded vertical bars provide the results of the VDA for STA, SO and Earth, respectively.
The middle panel combines the dynamic radio spectra from SWAVES observations with soft X-ray (SXR) measurements from GOES and  and the CME-heights obtained from LASCO-CDAW.
The gray-scale spectrogram displays radio emission intensity as a function of frequency and time (see Sect/ \ref{subsect:radio_burst} for more).
Overlaid in blue the GOES SXR time profile shows the C-class flare which peaks earlier than the inferred release time for Earth and SO but prior to the release time of the particles for STA.
The magenta dashed and dotted lines represent the linear and second order fit to the CME heights obtained from LASCO-CDAW.
Both illustrate the continuous expansion well after the flare maximum. 
The coloured vertical shaded regions and dashed lines identify the shock connectivity intervals associated with different observers, following the same colour scheme.
The lowest panel presents energetic particle intensity profiles measured by GOES and STA/HET.
The green curve corresponds to GOES 275-500 MeV protons, the red curve shows 60-100 MeV protons from STA/HET and the cyan curve depicts 63.1-75.11 MeV protons from SO/HET.
All SEP time profiles are time-shifted.
In particular, the highest STA/HET channel has an onset at 18:23 UT.
Assuming an apparent path length of 1.41 AU (as derived from the VDA), such particles would need $\sim$30 min to propagate from the Sun to STA. 
Hence, those would be released at $\sim$17:54 UT, which is consistent to the VDA of STA.
For GOES, similarly, the onset time of 275-500 MeV is marked on 18:20 UT, given the apparent path length of 0.88 AU these would propagate for $\sim$10 min and would have been released at 18:09 UT.
Finally, for SO, the 63.1-75.11 MeV protons would propagate for almost $\sim$41 min and their release would be marked at $\sim$17:54 UT.

\subsection{SEP composition}
\label{subsect:SEP abundances}

\begin{figure}[t]
\centering
\includegraphics[width=0.9\columnwidth]{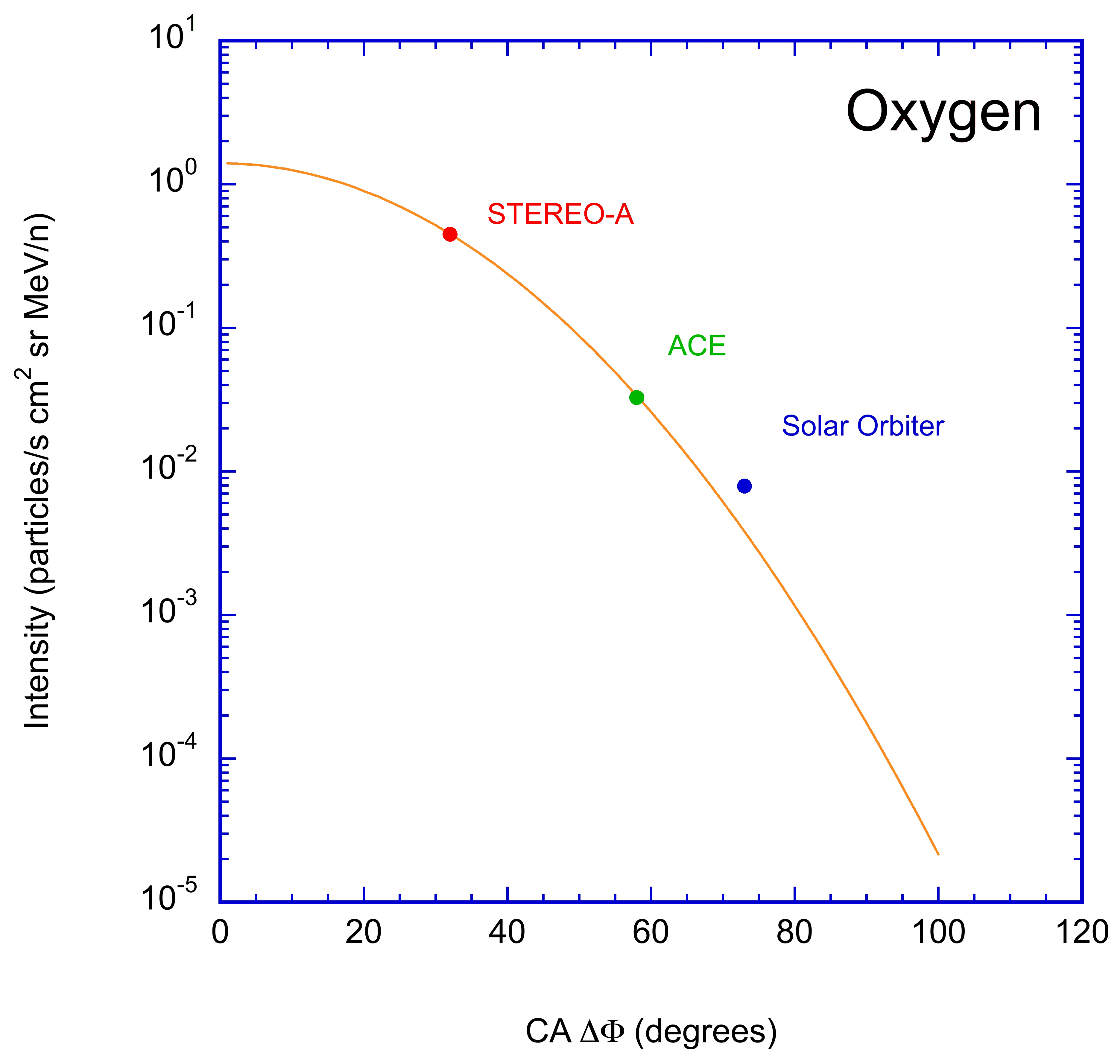}
\caption{Gaussian fit to the event-averaged 1.0 MeV/nucleon Oxygen intensities at each spacecraft versus distance to the spacecraft ($\Delta \phi$ in Table \ref{tab:fps_and_flare_locations}).
The sigma of the distribution of 21.2$\deg$}  
\label{fig:gaussian}
\end{figure}

The ion composition shows Fe-enrichment over a broad range of longitude connections in this event.
The STA Suprathermal Ion Spectrograph \citep[SIT;][]{Mason_2008}, Advanced Composition Explorer (ACE)/Ultra-Low Energy Isotope Spectrometer \citep[ULEIS;][]{Mason_1998} near Earth and SO/EPD-SIS \citep{Rodriguez-Pacheco_2020} show a clear enhancement of Fe during the SEP interval, together with $^{3}$He, $^{4}$He, O, and other heavy ions (See Fig. \ref{fig:composition}).
Such Fe enhancement has been observed in other widespread events, including GLE events \citep[][]{Cohen_2014, Cohen_2025}. 
\citet{Mewaldt_2012} surveyed GLE events during solar cycle 23 and found an average Fe/O ratio $\pm$twice that of gradual SEP events.  
These enhancement in the spectra in Fig. \ref{fig:composition}, however, are significantly higher, with Fe exceeding O at most energies.
There are trace amounts of $^{3}$He ($\pm$1\% or less) in these events, but other $^{3}$He-rich activity preceding the GLE makes it ambiguous as to whether or not the $^{3}$He is associated with the studied event.
The spectra in Fig. \ref{fig:composition} show a large variation in intensity between the different spacecraft, falling off as distance to the flare site increases.
Fig. \ref{fig:gaussian} shows a Gaussian fit in the manner of \citet{Cohen_2017} using the angular separations from Table \ref{tab:fps_and_flare_locations}.
The fitted Gaussian sigma is 21.2$^\circ$, similar to that found in other surveys of flares seen from multiple spacecraft \citep[][]{Cohen_2017}.   

\subsection{Radio bursts analysis}
\label{subsect:radio_burst}

Radio observations provide important diagnostics of particle acceleration and shock formation during eruptive solar events \citep[e.g. ][]{Kahler_1992, Cane_2002}.
In particular, type II radio bursts are commonly interpreted as signatures of electron acceleration at magnetohydrodynamic shocks propagating through the corona and interplanetary medium, and are therefore frequently associated with CME-driven shocks \citep[][]{Klein_1999, Warmuth_2015}.
Their spectral drift towards lower frequencies reflects the propagation of the emitting source into regions of decreasing plasma density, making them a valuable probe of the shock evolution and its relationship to SEP acceleration.
Multi-instrument observations provide a more complete view of the radio activity, from metric frequencies to the decametric-hectometric range (see Fig.~\ref{fig:radio-burst} in Appendix~\ref{annex: radio_spectra}).
Ground-based observations were limited, with only one CALLISTO station clearly detecting the radio activity associated with the eruption.
ROSWELL-NM recorded a metric type II burst between approximately 18:00 and 18:08 UT, with several distinct downward-drifting features observed between 250 and 400 MHz, possibly indicating electron acceleration at different regions of the expanding shock.
No type II burst was reported in the NOAA SWPC event lists.
However, the event was observed by the Owens Valley Radio Observatory Long Wavelength Array (OVRO-LWA) until approximately 18:25 UT, providing both dynamic spectra and radio imaging observations during the early phase of the event.
Early low-frequency activity appears in OVRO-LWA around 17:58--18:00 UT, coinciding with the rising phase of the hard X-ray emission observed by SO/STIX.
This early activity may be associated with flare-accelerated electrons and type III emission, although its identification remains uncertain.
The dynamic spectra show enhanced emission mainly around 20–40 MHz during this phase, followed by more complex structures.
A later type II burst was identified between 52 and 84 MHz.
STA/SWAVES also detected enhanced radio emission from around 18:00 UT, extending towards lower frequencies over the following hours.
The early broadband component may include type III emission, whereas the later, more prolonged component could be associated with the outward propagation of the CME-driven shock.
Faint type III emission was also reported by MEXICO-FCFM-UNACH and MEXICO-LANCE between approximately 18:02 and 18:03 UT \citep[][]{Monstein_2023}.
The onset of the metric type II emission, around 18:00 UT, is close to the reconstructed shock connection times to STA and Earth, estimated at approximately 17:58 and 18:01 UT, respectively.
This temporal correspondence is consistent with shock-associated electron acceleration during the early development of the magnetic connections to these observers.
Taken together, the observations are consistent with an early phase of flare-related electron acceleration followed by radio emission potentially associated with the expanding shock.
However, interference and the complexity of the observed spectral features prevent an unambiguous identification of all radio-burst components.

\begin{figure*}[t]
\centering
\includegraphics[width=0.65\columnwidth]{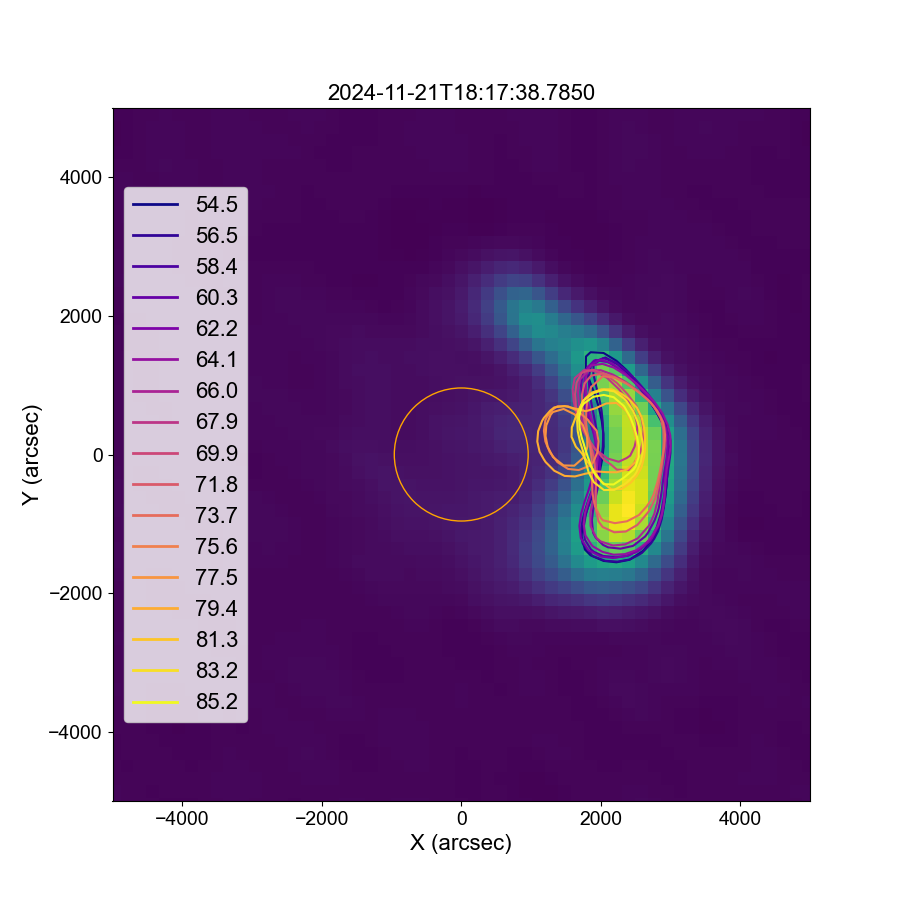}
\includegraphics[width=0.65\columnwidth]{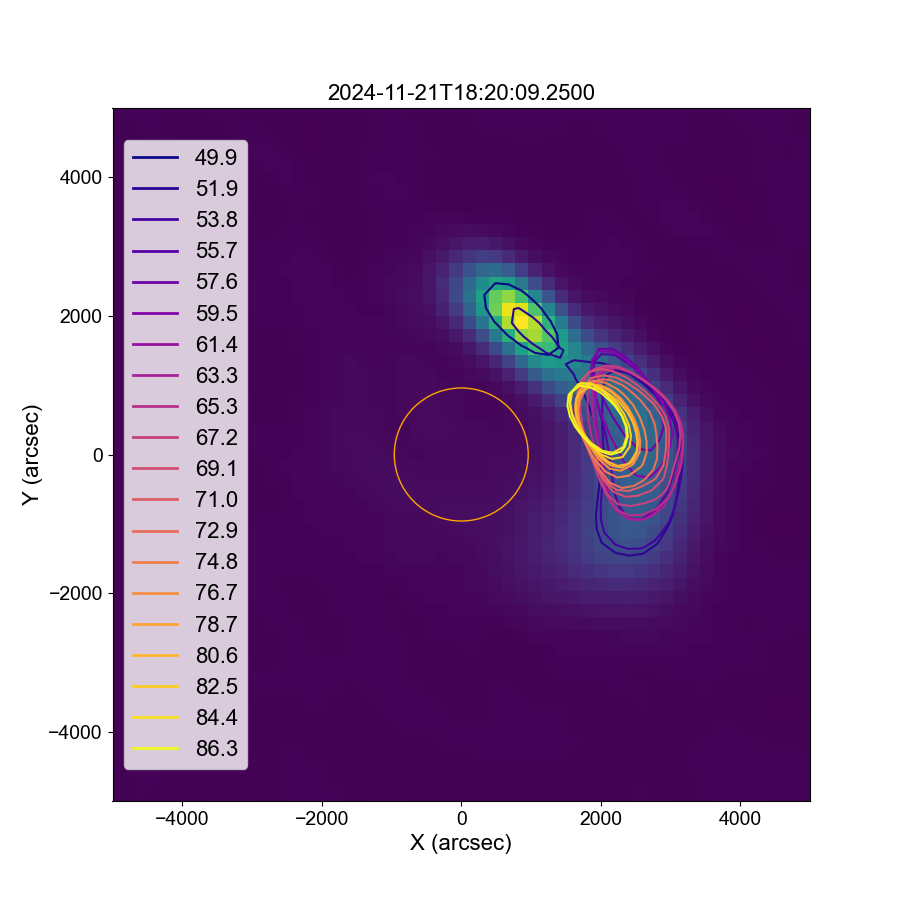}
\includegraphics[width=0.65\columnwidth]{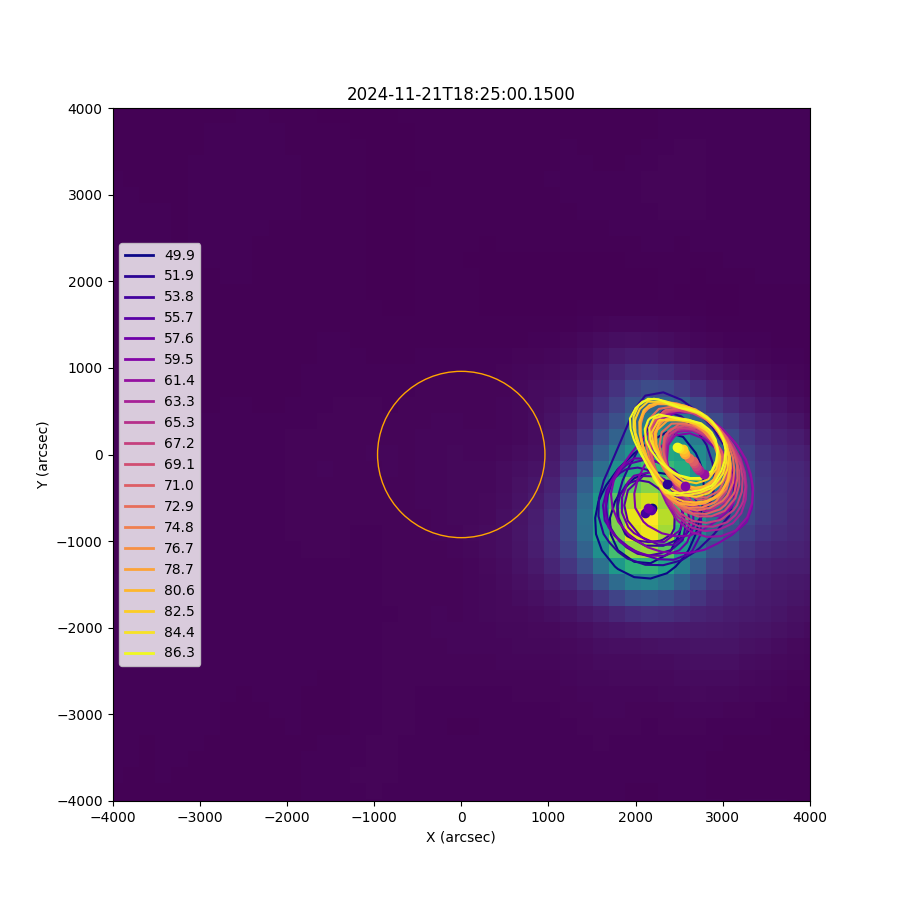}
\caption{Radio type II burst emission zone for three different timings, from left to right: 18:17, 18:20, and 18:25 UTC. The coloured contours represent different frequencies shown in the legend. The last image also shows the derived centroids of the radio sources.}  
\label{fig:2D_radio_centroids}
\end{figure*}

\begin{figure}[t]
\centering
\includegraphics[width=\columnwidth]{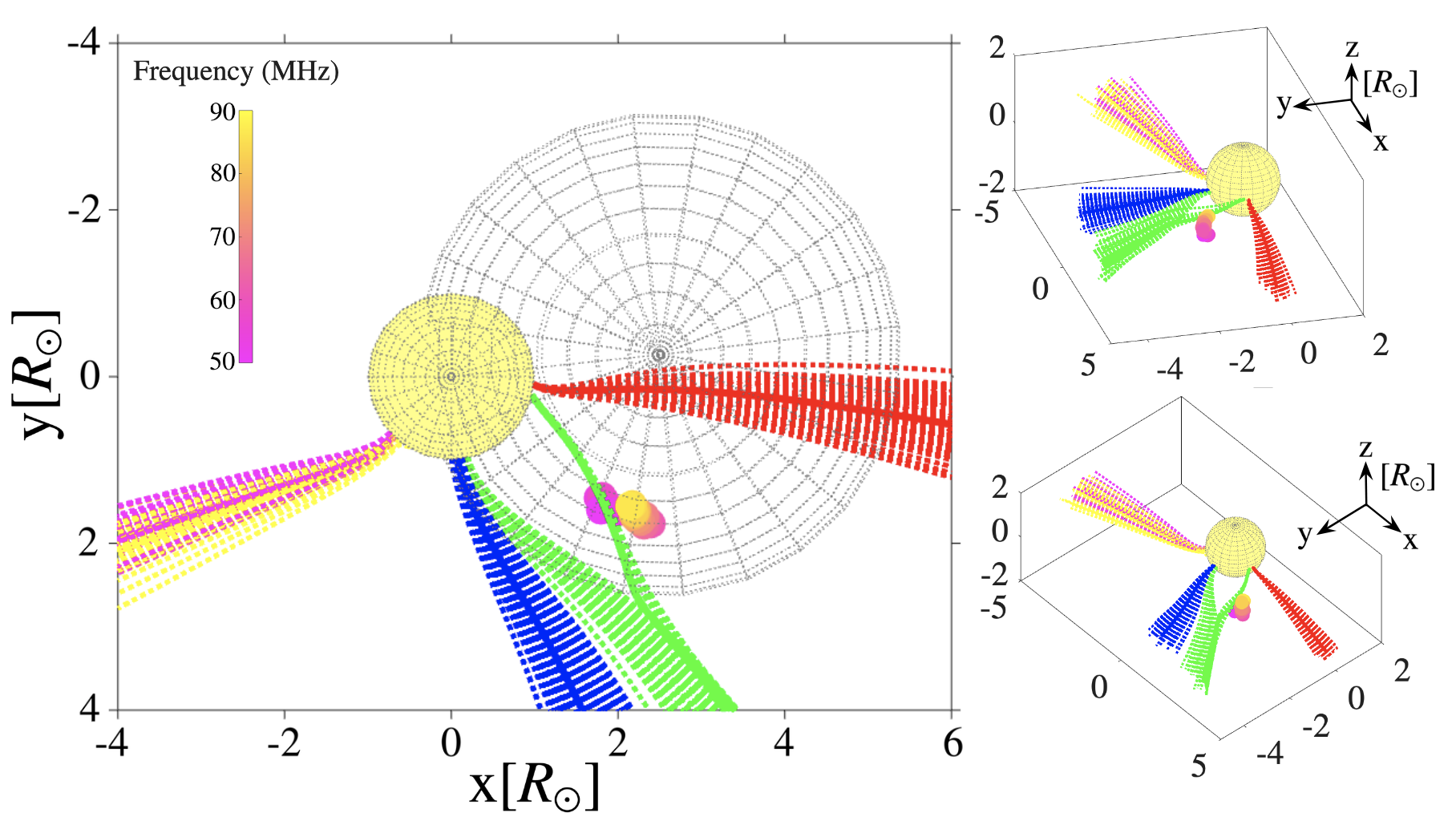}
\caption{Derived 3D centroids of the type II radio burst at 18:25 UTC (coloured bubbles) superposed to Fig. \ref{fig:3d_shock}. Left panel: view from the solar north, with the shock model represented by a gridded sphere. Right panels: two other views without the shock model. Bubbles colour from pink to yellow represents the frequency in MHz.}  
\label{fig:3D_radio_centroids_and_shock}
\end{figure}

OVRO-LWA observations provide an opportunity to investigate the spatial relationship between the radio-emitting regions and the CME-driven shock reconstructed from remote-sensing observations
Comparing the radio source locations with the reconstructed shock geometry provides additional constraints on the regions of the shock associated with radio emission during the early phase of the event.
The radio images were affected by ionospheric refraction, which introduced apparent positional shifts of the radio sources, particularly at lower frequencies.
To minimise these effects, only observations above 50 MHz were considered, as refraction becomes increasingly severe at lower frequencies.
In addition, only time intervals during which the quiet Sun was in its correct central position were selected, indicating limited ionospheric distortion.
The positions of the type II radio sources were identified in two-dimensional OVRO-LWA images at multiple frequencies (see Fig.~\ref{fig:2D_radio_centroids}).
The source centroids at 18:25 UT were selected for the 3D reconstruction because this time corresponds to the moment when the reconstructed magnetic connectivity is established between the shock and the three main observing spacecraft.
Their three-dimensional locations were inferred using the MAS MHD coronal density model, by associating the observed radio frequencies with the corresponding modelled coronal density surfaces \citep[using the methods presented in ][]{morosan2024}.
Figure~\ref{fig:3D_radio_centroids_and_shock} shows the resulting 3D type II radio source centroids in relation to the reconstructed shock and magnetic field lines.
Uncorrected ionospheric refraction may still affect the reconstructed source positions, potentially shifting the inferred locations away from the Sun.
The actual radio sources may therefore lie somewhat closer to the solar surface than indicated by the reconstruction.
This limitation is taken into account in the interpretation of the results and is further discussed in Section~\ref{sect:discuss_conclusion}.

\section{Discussion and Conclusion}
\label{sect:discuss_conclusion}

In this study, we performed a detailed analysis of the 21 November 2024 SEP event (GLE76).
This event provides an opportunity to investigate how the particle release, magnetic connectivity, and the evolution of a CME-driven shock combine to produce a relativistic SEP event detected by several spacecraft at different heliospheric locations.
The timing results indicate that both flare- and shock-related acceleration processes may have contributed to the relativistic SEP event in every observer.
The VDA-derived SRT at STA ($17{:}57\pm3$~min) and SO ($17{:}58\pm1$~min) coincide with the early flare phase, while the near-Earth release follows at $18{:}11\pm3$~min.
The timing differences indicates that the earliest particles sampled by the three observers cannot be described by a single release scenario under the assumptions of the VDA, considering that STA was the best magnetically connected observer, followed by Earth and SO.
This may reflect distinct injection histories, differences in access to the acceleration region, or propagation effects that influence the inferred release times and limitations on the VDA.

At STA, the nominal shock-connection time overlaps with the VDA release interval.
It also connects early to a strong shock regions that subsequently has substantially higher Mach number and compression ratio than any other shock regions connected to the other spacecraft.
These conditions are consistent with a shock contribution to the intense and prolonged SEP event observed at STA along with a contribution from a flare-related process.
The near-Earth release occurs after the nominal shock connection and after the subsequent development of stronger shock conditions, so the timing is compatible with a shock acceleration.
The Earth-connected shock regions also develop conditions favourable for shock-related acceleration, but the inferred parameters exhibit greater variability. 
The Type II radio observations (imaging) from OVRO-LWA show that there is an interaction of the shock with the HCS positioned near the location where Earth is also magnetically connected.
At this location conditions for very efficient SEP accelerated may develop at the shock.
The proximity of the connection to the HCS also allows for a contribution from a flare-related process and this is also confirmed further from the composition characteristics of the SEP event.
More specifically, we find the presence of $^3$He in ACE ULEIS observations suggesting that flare accelerated material had either direct access to the magnetic field lines connected to Earth or transported efficiently to the observer.
For SO, we find that its inferred particle release precedes its nominal connection to the reconstructed front by about 22 min.
Moreover, the subsequently connected regions exhibits comparatively weak shock conditions.
Therefore, its early particle release cannot be attributed to efficient acceleration at the locally connected shock.
The contrast with other observers shows that the SEPs measured in this location cannot be determined solely by the instantaneous properties of the shock at its nominal magnetic connection.
One possibility is that SEPs may retain signatures of earlier acceleration and transport or reach the observer from other portions of a more extended source that the model itself implies.
The SEP event composition from SIS observations also show presence of $^3$He suggesting also here that a flare-related process may have contributed in this location.

The combined observations and modelling show that GLE76 was associated with an accelerator whose accessibility and inferred efficiency varied between magnetic connections and observers.
We therefore favour a scenario in which an early particle population, plausibly including flare-accelerated particles, is supplemented by an evolving CME-shock contribution whose importance depends on the local acceleration properties and magnetic connectivity. 
The spectral evolution further supports this combined scenario.
STA and SO exhibit remarkably similar spectral softening and increasing particle intensities despite the very different shock conditions along their respective magnetic connections.
The observed spectra therefore do not simply follow the instantaneous efficiency of the locally connected shock.
At the same time, the increasing intensities and evolving high-energy rollover, particularly in the near-Earth measurements, are consistent with an increasing contribution from sustained CME-shock acceleration, with a possibility that the shock interaction with the streamer played an significant role to the observed properties in this location.
Curved spectral forms reproduce the high-energy spectra considerably better than a single power law, indicating an evolving accelerator together with energy-dependent transport.

We therefore interpret 21 November 2024 as the result of a flare-accelerated population whose signatures is depicted also in the composition of this event (Fe-rich and trace amounts of $^3$He), observed over a broad longitudinal range, onto which a shock-accelerated component is progressively superposed where the CME-driven shock develops favourable acceleration conditions at each location.
The relative importance of these components is controlled by the evolving shock properties, magnetic connectivity, and interplanetary particle transport. 
Quantifying the relative contributions will require coupled acceleration and transport modelling constrained by the particle timing, spectra, and radio observations. 
The principal result of this study is that the event cannot be understood through the properties of a single locally connected acceleration region alone.

\begin{acknowledgements}
This research received funding from the European Union’s Horizon Europe programme under grant agreement No 101135044 (SPEARHEAD) [\url{https://spearhead-he.eu/}]. Views and opinions expressed are however those of the author(s) only and do not necessarily reflect those of the European Union or the European Health and Digital Executive Agency (HaDEA). Neither the European Union nor the granting authority can be held responsible for them. SO is a space mission of international collaboration between ESA and NASA, operated by ESA. Solar Orbiter post-launch work at Johns Hopkins University Applied Physics Laboratory (JHU/APL) is supported by NASA contract NNN06AA01C. The authors acknowledge the OVRO-LWA expansion project, which was supported by NSF under grant AST-1828784. OVRO-LWA solar operations are supported by NSF grant AGS-2436999 to New Jersey Institute of Technology. A.K. acknowledges financial support from NASA LWS grant 80NSSC25K0130 and HGIO grant 80NSSC24K0555.
\end{acknowledgements}

\bibliographystyle{aa}
\bibliography{main.bib}

\appendix

\section{Additional material on the event}
\label{annex: additional_material}

\begin{figure}[h]
\centering
\includegraphics[width=0.45\columnwidth]{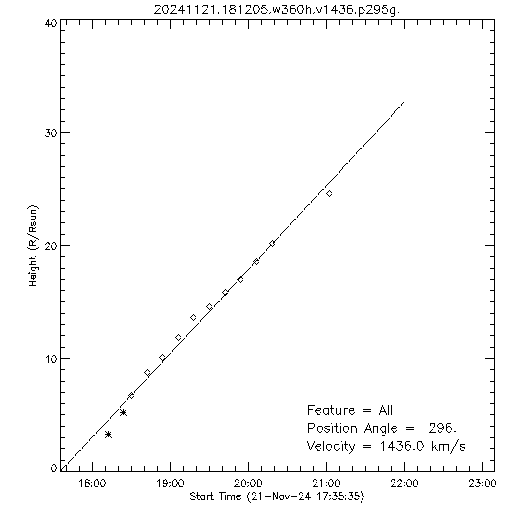}
\includegraphics[width=0.45\columnwidth]{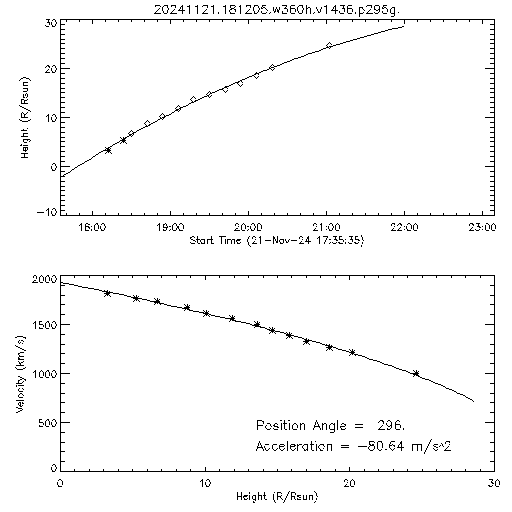}
\caption{Left panel: CME height [R$_\odot$] as a function of time in hours from SOHO/LASCO measurements for the 21 November 2024 event. The line represents the linear fit which yields a CME speed of 1436 km/s. Top right panel: same measurements on which a second order fit is derived. Bottom right panel: CME speed [km/s] as a function of height [R$_\odot$], which yield a CME acceleration of -80.64 m/s$^2$.}  
\label{fig:SOHO-LASCO_CME_fit}
\end{figure}

\begin{figure}[h]
\centering
\includegraphics[width=\columnwidth]{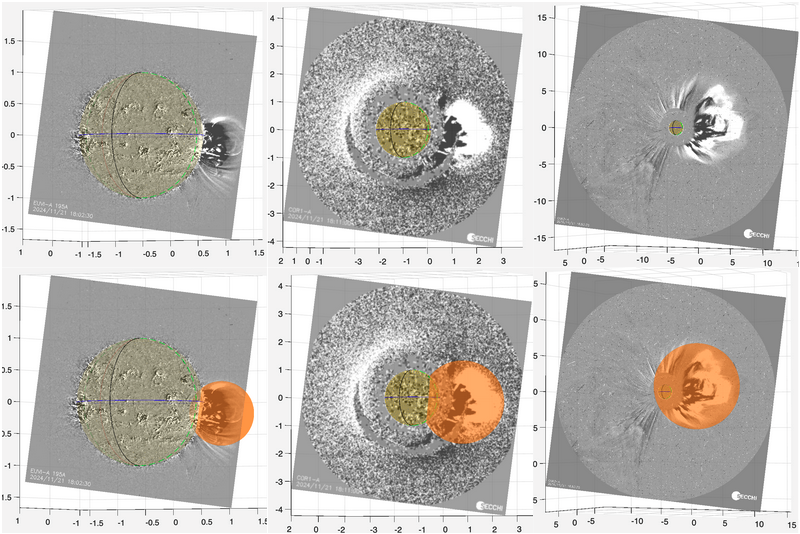}
\caption{Reconstructed 3D shock wave on EUV and white-light images. Top panels, from left to right: images of the expanding shock wave from STA with, respectively, EUVI at 18:02:30 UT, COR1 at 18:11:00 UT, and COR2 at 18:52:30 UT. Bottom panels: same images on which the reconstructed 3D shock wave, represented by an orange ellipsoid, is superposed.}  
\label{fig:shock_fit}
\end{figure}

\begin{figure*}[h]
\centering
\includegraphics[width=0.95\textwidth]{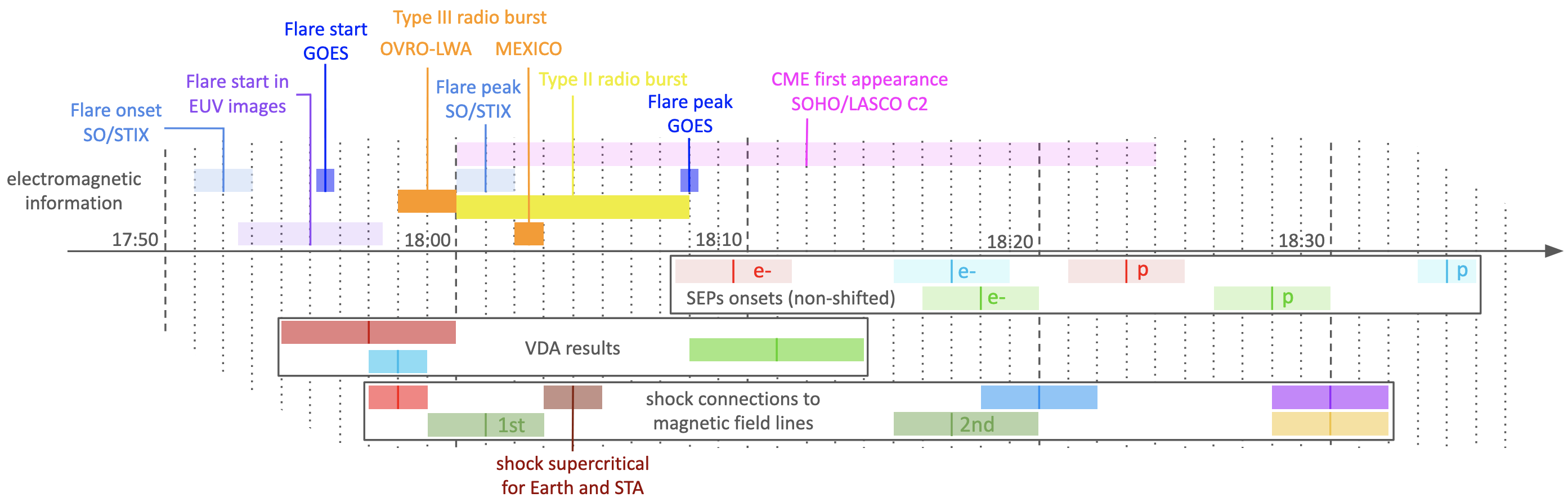}
\caption{Timeline of the event, at the observer time. Top: electromagnetic information (Sect. \ref{subsect:remote_obs}). The solid line indicates the timing, and the transparent band represents its uncertainty. Bottom: Red, green, cyan, purple and yellow bands represents, respectively, information on STA, Earth, SO, PSP and Bepi spacecraft. First bloc shows SEPs onsets for the three main spacecrafts (Sect. \ref{subsect:sep_obs}), second bloc displays the results of the VDA (Sect. \ref{subsect:sep_obs}), and third bloc notes the time of shock connection to magnetic field lines (Sect. \ref{subsect:MHD_parameters}.}  
\label{fig:timeline}
\end{figure*}

\section{Previous solar activity}
\label{annex: solar_activity}

As mentioned briefly in the Sect. \ref{subsect:remote_obs}, the Sun was very active on the 21 November 2024.
Three previous flares were recorded.
Table \ref{tab:previous_flares} list and compare them with the flare associated with the SEP event studied, using the GOES flare event list\footnote{\url{https://umbra.nascom.nasa.gov/goes/eventlists/goes_event_listings/}} and the SO/STIX science flare list\footnote{\url{https://github.com/hayesla/stix_flarelist_science/}}.

\begin{table*}[h]
\caption{Previous flare activity on the 21 November 2024}
    \centering
    \small
    \begin{tabular}{l|ccccc|ccccc}
     & \multicolumn{5}{c|}{GOES} & \multicolumn{5}{c}{SO/STIX}\\
    Name & Start & Peak & Stop & Intensity & Identified AR & Start & Peak & Stop & Intensity & Identified AR\\
    \hline
    XR1 & 12:45 & 13:07 & 13:24 & C3.4 & AR~13901 (S07E33) & 12:51 & 13:00 & 13:08 & 2687 & AR~13901 (S09E33)\\
    XR2 & 13:24 & 13:52 & 14:15 & C5.4 & AR~13901 (S07E33) & 13:44 & 13:47 & 13.55 & 11775 & AR~13901 (S09E31)\\
    XR3 & 14:50 & 14:57 & 15:10 & C4.7 & AR~13905 (S10E89) & 14:50 & 14:55 & 15:04 & 5375 & AR~13905 (S13E93)\\
    Main event & 17:52 & 18:07 & 18:28 & C5.9 & AR~13901 (S08E31) & 17:58 & 18:01 & 18:07 & 15871 & AR~13901 (S09E30)\\
\end{tabular}
\tablefoot{SO/STIX intensity shows the 4-10~keV peak counts.}
\label{tab:previous_flares}
\end{table*}

\begin{figure}[h]
\centering
\includegraphics[width=0.8\columnwidth]{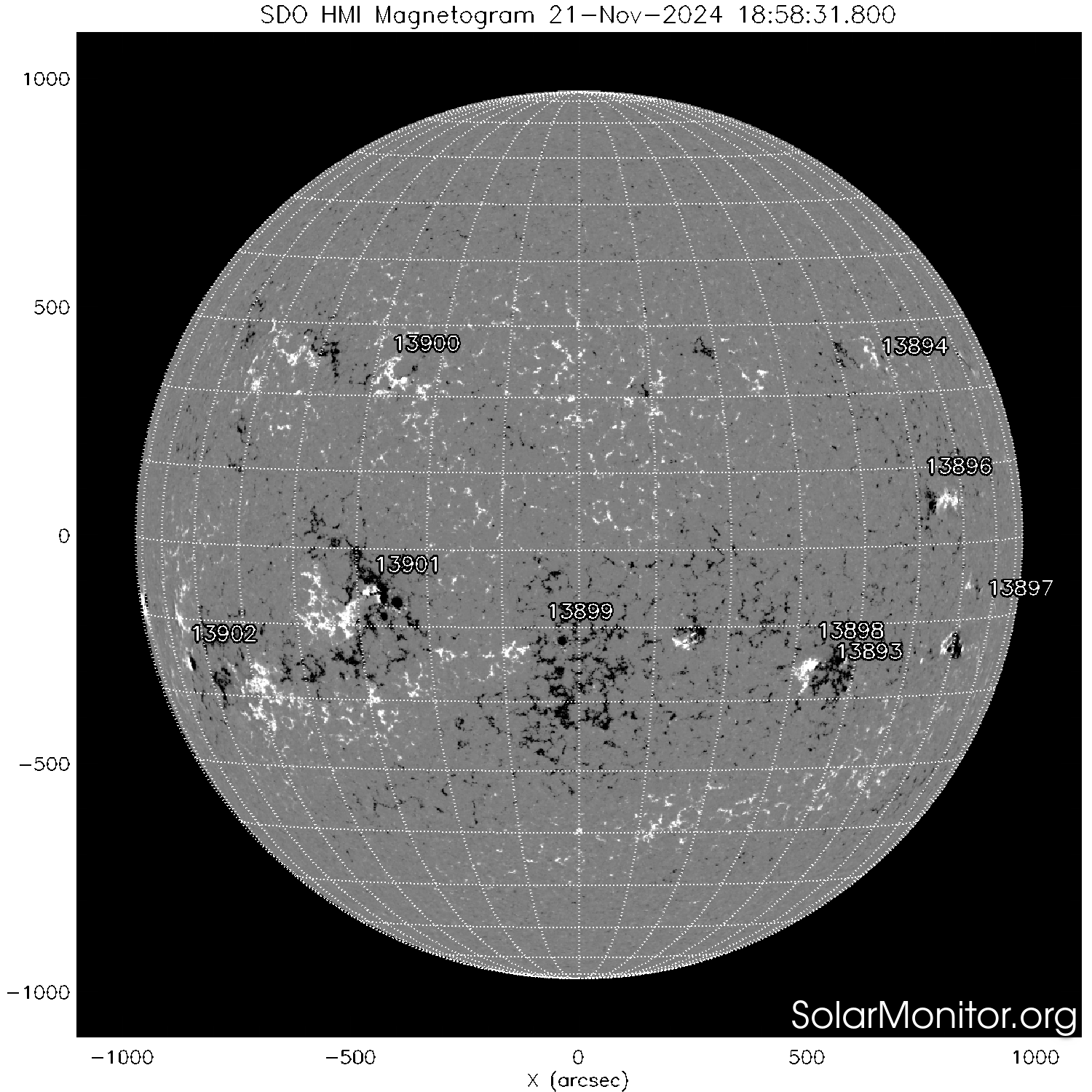}
\caption{SDO/HMI magnetogram of the Sun on the 21 November 2024 at 18:58:31 UT, on which numerous active regions are visible.}
\label{fig:sun_ARs}
\end{figure}

The Sun also presents other active zones (see Fig.~\ref{fig:sun_ARs}), with sudden increase of brightness in EUV and visible images, like the non-enumerated AR close to S07W64 (CRLN $\sim 312$), and the zone including the closed AR~13893 and AR~13898 ($\sim$ S17W40).
A movie from SDO/AIA showing several confined flares in these area is available in supplementary material.
This zone that are very well magnetically connected to SO spacecraft, but also to the second packet of Earth field lines footpoints (see Section \ref{subsect:mag_connectivity}).

\section{Radio emission spectra}
\label{annex: radio_spectra}

\begin{figure*}[h]
\centering
\includegraphics[width=0.8\textwidth]{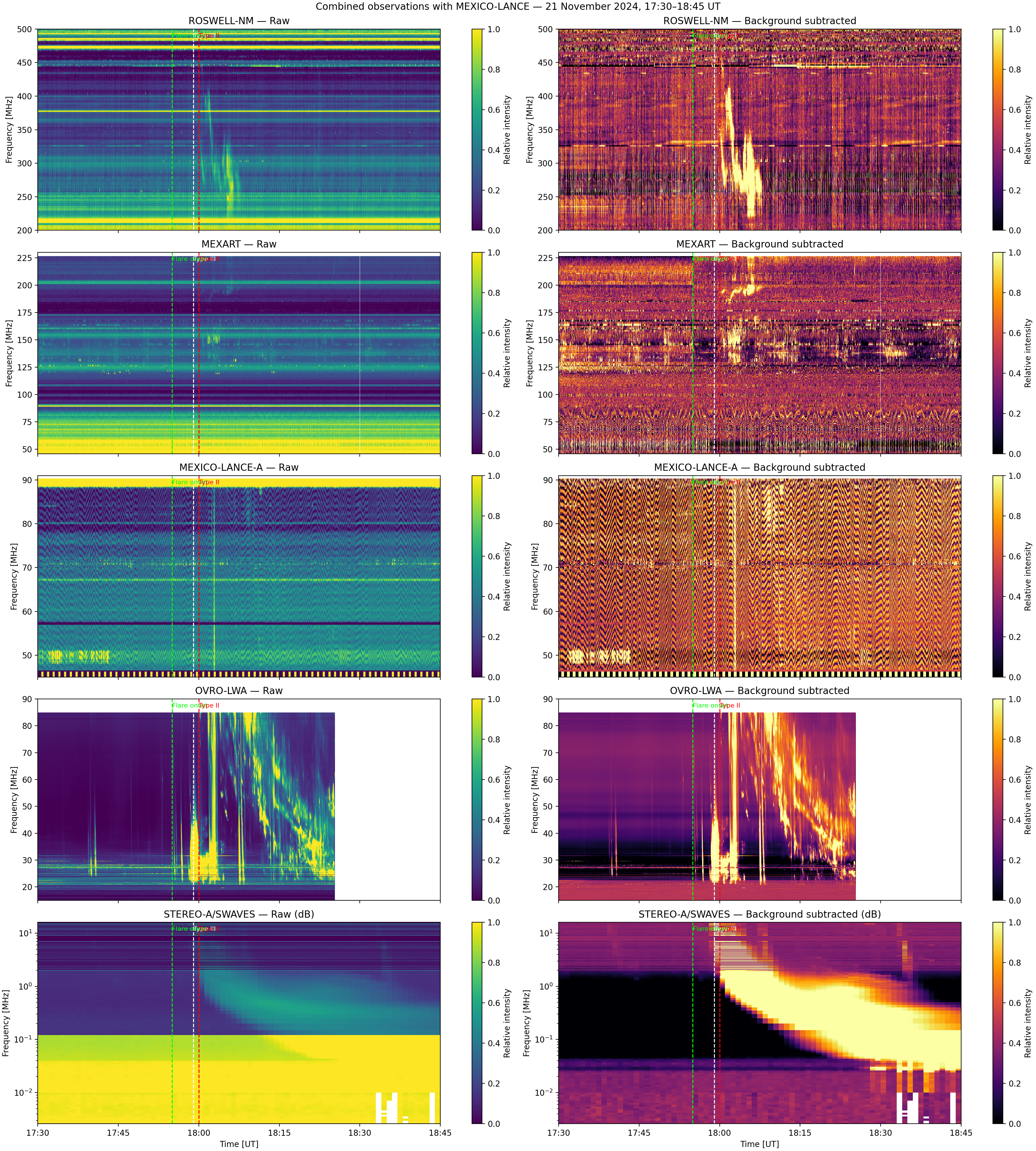}
\caption{Multi-instrument radio observations of the 21 November 2024 event from 17:30 to 18:45 UT. The dynamic spectra combine ground-based observations from (from top to bottom) ROSWELL-NM, MEXART, MEXICO-LANCE and OVRO-LWA with space-based observations from STA/SWAVES. Left panels shows their raw spectra and right panels shows background-subtracted spectra, coloured according to their relative intensity. On each spectrum, vertical dashed lines represents the flare onset in hard X-ray (green), the type III burst onset (white), and the type II burst (red).}  
\label{fig:radio-burst}
\end{figure*}

\end{document}